\documentclass[fleqn,usenatbib]{mnras}
\usepackage{hyperref}
\usepackage[T1]{fontenc}
\usepackage{ae,aecompl}
\usepackage{graphicx}
\usepackage{amsmath}
\usepackage{amssymb}
\usepackage{pdflscape} 
\usepackage{footnote}
\usepackage{threeparttable}
\usepackage{epsfig}
\usepackage{pslatex}
\usepackage{amsmath}
\usepackage{epstopdf}
\usepackage{csquotes}
\usepackage{orcidlink}
\graphicspath{{./}{figures/}}
\title[RS Oph 2021]{Study of 2021 outburst of the recurrent nova RS Ophiuchi: Photoionization and morpho-kinematic modelling}
\author[Pandey et al.]{
Ruchi Pandey\,\orcidlink{0000-0002-6222-3045}$^{1}$\thanks{E-mail: ruchi.pandey@bose.res.in},
Gesesew R. Habtie\,\orcidlink{0000-0001-9827-738X},$^{1}$
Rahul Bandyopadhyay\,\orcidlink{0000-0001-9161-0397},$^{1}$
Ramkrishna Das\,\orcidlink{0000-0002-5440-7186},$^{1}$
\newauthor
Fran\c{c}ois Teyssier,$^{2}$
Joan Guarro Flo,$^{2}$
\\
$^{1}$S N Bose National Centre for Basic Sciences, Salt Lake, Kolkata 700 106, India\\
$^{2}$ ARAS Eruptive Stars Group\\ 
}

\date{Accepted XXX. Received YYY; in original form ZZZ}

\pubyear{2022}

\begin{document}
\label{firstpage}
\pagerange{\pageref{firstpage}--\pageref{lastpage}}
\maketitle
\begin{abstract}
We present the evolution of the optical spectra of the 2021 outburst of RS Ophiuchi (RS Oph) over about a month after the outburst. The spectral evolution is similar to the previous outbursts. Early spectra show prominent P Cygni profiles of hydrogen Balmer, \ion{Fe}{ii}, and \ion{He}{i} lines. The emission lines were very broad during the initial days, which later became narrower and sharper as the nova evolved. This is interpreted as the expanding shocked material into the winds of the red giant companion. We find that the nova ejecta expanded freely for $\sim 4$ days, and afterward, the shock velocity decreased monotonically with time as $v\propto t^{-0.6}$. The physical and chemical parameters associated with the system are derived using the photoionization code \textsc{cloudy}. The best-fit \textsc{cloudy} model shows the presence of a hot central white dwarf source with a roughly constant luminosity of $\sim$1.00 $\times$ 10$^{37}$ erg s$^{-1}$. The best-fit photoionization models yield absolute abundance values by number, relative to solar of He/H $\sim 1.4 - 1.9$, N/H = $70 - 95$, O/H = $0.60 - 2.60$, and Fe/H $\sim 1.0 - 1.9$ for the ejecta during the first month after the outburst. Nitrogen is found to be heavily overabundant in the ejecta. The ejected hydrogen shell mass of the system is estimated to be in the range of $3.54 - 3.83 \times 10^{-6} M_{\odot}$. The 3D morpho-kinematic modelling shows a bipolar morphology and an inclination angle of $i=30^{\circ}$ for the RS Oph binary system. 
\end{abstract}
\begin{keywords}
stars : novae, cataclysmic variables ; stars : individual (RS Ophiuchi)
\end{keywords}

\section{Introduction}\label{sect:intro}
RS Ophiuchi (RS Oph) is a well-studied galactic recurrent nova and symbiotic binary system with an average recurrence timescale of about 15 years \citep{2010Schaeffer}, and an orbital period of 453.6 days \citep{2009brandi}. The system has undergone six previous recorded outbursts in the years 1898, 1933, 1958, 1967, 1985, and 2006 \citep{2017ApJ...847...99M, 2010Schaeffer}. The system is also suspected to have had outbursts in 1907 and 1945 when it was aligned with the sun \citep{1993AAS...183.5503O, 2004Schaefer}. The RS Oph binary system has a red giant companion with a mass of $0.68 - 0.80 M_{\odot}$ \citep{2009brandi} and the spectral type of M2$-$M3 \citep{2018Mondal}. 

The previous outbursts of RS Oph in 2006 have been studied intensively from radio to X-rays. The X-ray observations showed that the emission from the shocked material formed when the nova ejecta with very high velocity interacted with a dense circumstellar medium \citep{2006BodeRSOphUV}. In a recent study, \citet{2022ApJMontez} studied the X-ray-emitting plasma moving outward from the WD using X-ray data observed from Chandra in the years 2007, 2009, and 2011. They detected jet-like structures spreading out from the system to the east and west, mirroring the structures seen at other wavelengths. \citet{Iijima09} detected a short lived flare-up of the \ion{He}{i} emission line in their optical observations during 2006 outburst, which they attributed to a helium flash. \citet{2006DasRSOph} investigated near-infrared J band ($1.0 - 1.35 ~\mu\text{m}$) spectra during 2006 outburst and detected an infrared shock wave as the nova ejecta passed through the red giant winds. \citet{2007MNRASEvans} studied the early phase of the spectral evolution of RS Oph in Infrared (IR), covering the wavelength range $0.87-2.51\mu$m. The radio observations revealed the bipolar outflows in RS Oph taken between 14 and 93 days after the outburst \citep{2006NaturO'brien}. The 1985 outburst of RS Oph was the first to be widely observed over the whole electromagnetic spectrum, from radio to X-ray. The presence of shock in RS Oph was first detected during the 1985 outburst in the form of bright radio synchrotron emissions \citep{1986ApJ...305L..71H}. It was extensively observed by the IUE satellite, both at low and at high resolution in the range of 1200 - 3000 {\AA}, during its 1985 outburst \citep{1996Shore}. The spectral evolution of the 1985 outburst was studied in optical and IR by \citet{1987rosino}. The first in-depth study of the time evolution of RS Oph in $1 - 3.5 ~\mu$m for about 3 years after the outburst is reported by \cite{1988EvansIR}. \cite{1969CoKon..65..257B} reported the outburst in 1967. A review of the historical outbursts was reported by \citep{1987rorn.conf....1R}. A recent study by \cite{2017ApJ...847...99M} indicates that the RS Oph system has a CO-type white dwarf (WD) with a mass expected to be in the range of $1.2 - 1.4 M_{\odot}$. The WD of the RS Oph system has most likely been increasing in mass due to the accumulation of a percentage of matter on its surface, and it may eventually approach the Chandrasekhar limit and explode as a Type Ia supernova \citep{2016Booth, 2008A&AWalder, 2001ApJHachisu}.

The most recent recorded seventh outburst of RS Oph was detected on 2021 August 8.93 UT (JD 2459435.430) by K. Geary (VSNET alert N.26131\footnote{\url{http://ooruri.kusastro.kyoto-u.ac.jp/mailarchive/vsnet-alert/26131}}) with an estimated visual magnitude of 5.0. The current 2021 outburst of RS Oph is also being monitored over a broad electromagnetic spectrum, from gamma rays to the radio domain \citep{2021ATelFermi-LAT, 2021ATelMAXI, 2021ATelSwift-XRT, 2021ATelNicer, 2021ATelSALT, 2021ATelINTEGRAL, 2021ATelARAS, 2021ATelpolarisation, 2021ATelIR, 2021ATelAstrosat}. RS Oph reached maximum brightness on Aug 10.1 with a V band magnitude of 4.5 and entered the supersoft source (SSS) phase on Sep 4 \citep{2021ATelPage, 2021ATelorio}. The spectral development of the present outburst of RS Oph is also similar to the spectra obtained during previous outbursts. A pictorial atlas of the spectroscopic evolution of RS Oph for the first 18 days, and from day 19 to 102 is presented in \cite{2021Munari}, and \cite{2022arXivMunari}, respectively.

In the present paper, we study the evolution of optical spectra of the 2021 outburst for over a month following the outburst using data from the ARAS database (see section~\ref{sect:data} for details). We used the photoionization code \textsc{cloudy}, v.17.02 \citep{2017RMxAA..53..385F} to analyse the observed spectra and construct a simple phenomenological model based on a spherical geometry to estimate the physical and chemical parameters of the system during 2021 outburst. To obtain the ejecta morphology and inclination angle of the binary system, we modelled the prominent \ion{H}{$\alpha$} emission line using 3D morpho-kinematic modelling and reconstruction tool, \textsc{shape} \citep{2011Shape}. The paper is organised as follows. In Section~\ref{sect:data}, we discuss the details of observations. The results and discussions are presented in Section~\ref{sect:results} in detail. In Section~\ref{sect:summ}, we present the summary.

\section{Data set}\label{sect:data}
We used publicly accessible medium and high-resolution optical spectra from the Astronomical Ring for Access to Spectroscopy Database (ARAS Database\footnote{\url{https://aras-database.github.io/database/index.html}}; \cite{2019CoSka..49..217T}) for the present study. The ARAS initiative consists of independent, small telescopes (from 20 to 60 cm), equipped with spectrographs  (resolution, $R$ $\sim$ 500 to 15000), which cover the wavelength range from 3600 to nearly 9000 {\AA}. The early spectral evolution from Aug 9.8 was intensively monitored by the ARAS group for over 3 months \citep{2021ATelARAS}. We selected a total of 18 optical spectra ($R > 9000$) for the present study. The log of observations is presented in Table \ref{tab:log}. Spectra taken by Fran\c{c}ois Teyssier were obtained with an Echelle spectrograph Resolution 9000 with CCD ATIK 460 EX mounted on a Schmidt-Cassegrain 16" telescope in the Santa Maria de Montmagastrell observatory, while spectra by Joan Guarro Flo used the same setup in Pierra (Spain). Furthermore, the spectra taken by Olivier Thizy were obtained with a spectrograph eShel (Shelyak) Resolution 11000 with CCD ATIK 460 EX mounted on a Schmidt-Cassegrain 11" telescope in the Observatoire de la Belle Etoile, Revel (France).

The spectra were reduced by following standard procedures using Integrated Spectroscopic Innovative Software (ISIS\footnote{\url{http://www.astrosurf.com/buil/isis-software.html}}). All spectra were then de-reddened with E(B - V) = 0.73 \citep{1987snijders}. The spectra presented here are not flux calibrated. We normalized all of the spectra to \ion{H}{$\beta$} line (see Figure~\ref{fig:spectralevolution}).
Flux and widths of emission lines were measured interactively using the tasks available in \textsc{iraf}\footnote{\textsc{iraf} is a package of software developed by the National Optical Astronomy Observatory for the reduction of astronomical images in pixel array form. \url{http://ast.noao.edu/data/software.}}.
\begin{table}
\centering
\caption[]{Journal of observations of RS Oph during 2021.\label{tab:log}}
\setlength{\tabcolsep}{6pt}
\small
\begin{threeparttable}
\centering
\begin{tabular}{lclcl}
\hline\noalign{\smallskip}
Date (UT) & Days after outburst & Wavelength ({\AA}) & Observer \\
\hline
Aug 9.88 & 0.95   & 3944-8942 & 1\\
Aug 9.99 & 1.06   & 3944-8942 &  1\\
Aug 11.87 & 2.94   &4197-7361  & 2\\
Aug 12.83 & 3.91    &3874-8941  & 3\\
Aug 13.87 & 4.94  & 4000-8900  & 1\\
Aug 16.83 & 7.90   & 3944-8940  & 1\\
Aug 17.82 & 8.89  & 3950-8899  & 1\\
Aug 18.82 & 9.89   & 3874-8940 & 1\\
Aug 19.85 & 10.92  & 4200-8899  & 1\\
Aug 20.89 & 11.96   & 3874-8939  & 1\\
Aug 21.91 & 12.98  & 3874-8941  &1 \\
Aug 22.91 & 13.98   &3874-8942  & 3\\
Aug 23.86 & 14.93   & 4200-8899  & 1\\
Aug 25.82 & 16.89   & 3874-8941  & 1\\
Aug 26.81 & 17.88   & 3874-8939 & 1\\
Aug 28.80 & 19.87    &3807-8942  & 3\\
Aug 29.80 & 20.87   &3807-8942  & 3\\
Aug 30.87 & 21.94   &3807-8943  & 3\\
\hline
\end{tabular}
\begin{tablenotes}
\item Observers: (1) Fran\c{c}ois Teyssier (FTE), (2) Olivier Thizy (OTH), (3) Joan Guarro Flo (JGF)
\end{tablenotes}
\end{threeparttable}
\end{table}

\section{Results and discussion}\label{sect:results}
\subsection{Light curve}
\begin{figure}
\centering
\includegraphics[scale=0.68]{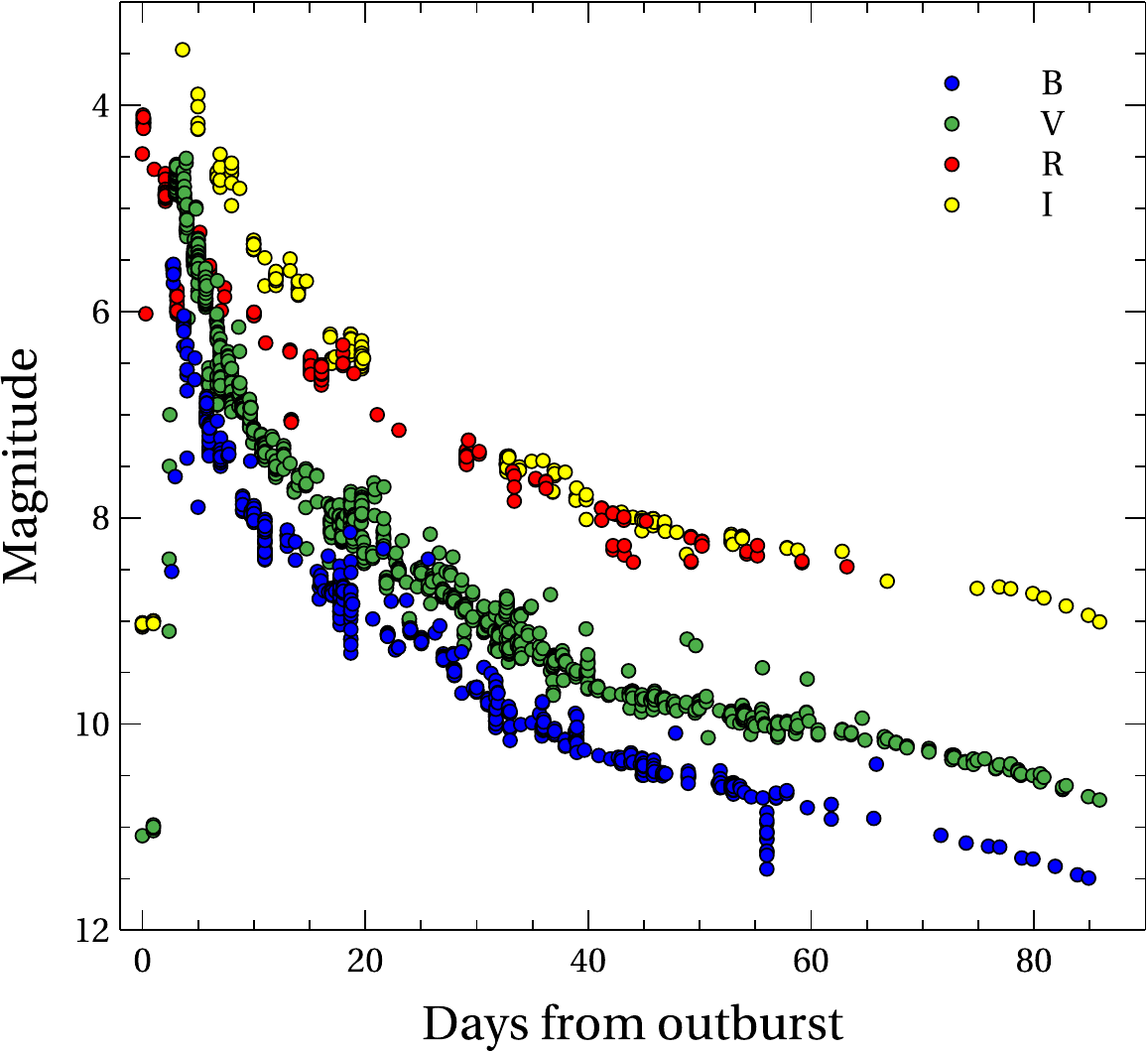}
\caption{The evolution of the \textit{B} (blue circle), \textit{V} (green circle), \textit{R} (red circle), and \textit{I} (yellow circle) light curve for RS Oph during the first 3 months after the 2021 outburst. \label{fig:lightcurve}}
\end{figure}

The evolution of the \textit{BVRI} light curve for RS Oph during the first 3 months after the 2021 outburst is presented in Figure~\ref{fig:lightcurve}. The data are taken from the American Association of Variable Star Observers (AAVSO\footnote{\url{https://www.aavso.org/}}) database. The nova was first detected on 2021 August 8.93 UT, 22:20, with an estimated V band magnitude of 5.0, which we considered as the day of the outburst (+0 day) in the present paper. The nova reached maximum brightness on Aug 10.1 with the V band magnitude of 4.5 and the B band magnitude of 5.5. The optical brightness of a nova declined after reaching its maximum. The decline in the light curve has been observed to be smooth and similar to the previously recorded outbursts. The decline time scales $t_2$ and $t_3$ are determined to be 2.77 days and 8.04 days, respectively, in the V band, and 4.23 days and 12.90 days, respectively, in the B band. The decline time scales during the 2006 outburst in B band were $t_2=6.2$ days and $t_3=17.1$days \citep{2007BaltA..16...46M}. For previous recorded outbursts, \cite{2010Schaeffer} reported $t_3 = 14$ days for RS Oph. For the outburst in the years 1958 and 1933, \cite{1985cassatella} determined the decline time scales $t_2$, and $t_3$ to be 3 and 7 days, respectively. The distance of RS Oph from maximum absolute magnitude relation with decline time (MMRD; \cite{1995MMRD}) is estimated to be $1.68$ kpc by assuming E(B-V)=0.73 $\pm$ 0.10 \citep{1987snijders}. In a multi-frequency observation of the 1985 outburst of RS Oph, \cite{1987snijders} calculated the reddening in the direction of RS Oph by removing the interstellar extinction feature at 2175 {\AA} in IUE spectra, giving E(B-V)=$0.73 \pm 0.10$.
The study of the previous outburst of RS Oph in 1985 in the multi-wavelength domain estimated a distance, $d = 1.6 \pm 0.3$ kpc \citep{1987Bode}. For the 2006 outburst, \cite{2008ASPC..401...52B} estimated a distance of RS Oph to be $1.4^{+0.6}_{-0.2}$ kpc. \cite{2006Sokoloski} also reported a distance of $\sim 1.6$ kpc for RS Oph. However, distances to the RS Oph are still a subject of debate. \citet{1987snijders} reviewed several distance determination methods and concluded that 1.3$< D(kpc) <$ 2.0, with a most probable value of 1.6kpc. A historical record on the distance determination of RS Oph is reported by \cite{2008ASPC..401...52B}.

\subsection{Optical spectra}
\begin{figure*}
\centering
\includegraphics[scale=0.63]{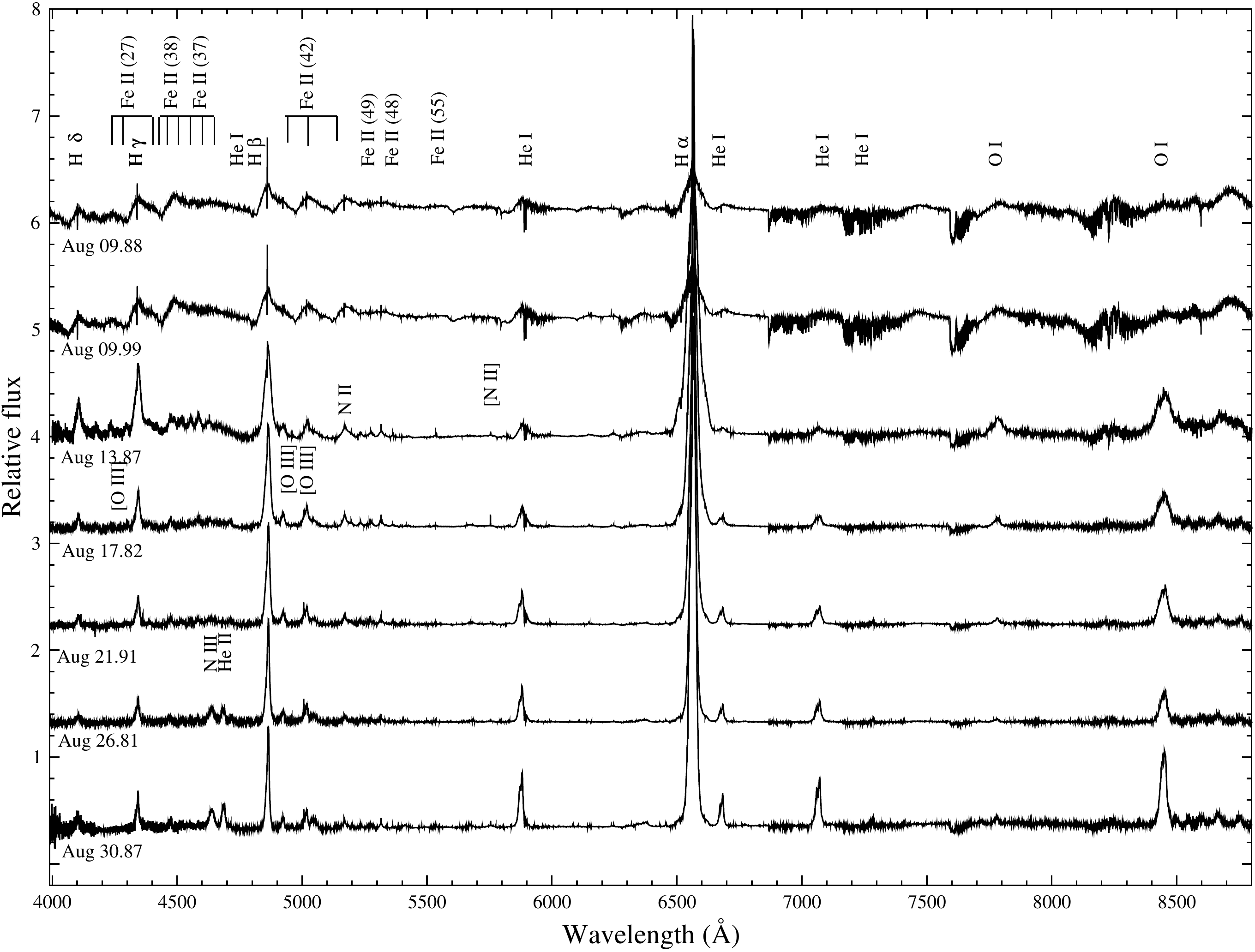}
\caption{Evolution of optical spectra of RS Oph during 2021 outburst. The spectra have been arbitrarily scaled and all the prominent emission lines are labeled. The date of outburst, 2021 August 8.93 UT is considered as $t=0$. The spectra presented here are not corrected for reddening. \label{fig:spectralevolution}}
\end{figure*}

The evolution of optical spectra for RS Oph during the first months after the 2021 outburst is presented in Figure~\ref{fig:spectralevolution}. The spectra cover the region from 4000 {\AA} to 9000 {\AA}. The prominent emission lines are marked in the figure. The spectral evolution during the 2021 outburst is similar to that observed during the 2006 outburst.

The spectrum of nova passes through different stages \citep{2012BASIShore}. The first stage of expansion is an optically thick fireball phase \citep{1998Gehrz}. The spectrum of RS Oph just one day after an outburst during the fireball phase was dominated by broad P Cyg profiles of hydrogen Balmer lines, \ion{Fe}{ii}, \ion{O}{i}, along with the weaker and broad lines of \ion{He}{i} (4924, 5876, 6678, 7065 {\AA}) and \ion{O}{i}. The ionisation levels of the ejecta were low during the initial stages, and the nova spectrum was usually dominated by the permitted recombination lines. As the spectrum evolved to its optical maximum, the non-Balmer lines of \ion{Fe}{ii} and \ion{O}{i} 7774 {\AA}$(3s^5S^0-3p^5P)$ and 8446 {\AA}$(3s^3S^0-3p^3P)$ started becoming prominent. The emission line of \ion{O}{i} 7774 {\AA} arises primarily from the recombination \citep{1991ApJRudy}. The emission lines of \ion{Fe}{ii} multiplets in the spectrum of RS Oph were prominent from 3 days after the outburst date. This stage is known as the Fe-curtain phase \citep{2016A&AShore}. Out of all \ion{Fe}{ii} multiplets present in the spectrum, the \ion{Fe}{ii} (42) showed the most prominent emission features at 4924 {\AA}, 5018 {\AA}, and 5169 {\AA}. The forbidden line [\ion{N}{ii}] 5755 {\AA} was first appeared around August 10, with only a narrow component, similar to the 2006 outburst \citep{Iijima09}. Its intensity increased after that and had only a narrow component. This emission line most probably had its origin in the ionised winds of the red giant companion \citep{2018Mondal}. The continuum flux decreased after August 10, and the fluxes of \ion{O}{i} and \ion{He}{i} increased. The emission line of \ion{He}{i} 5876 {\AA} became more prominent.

The P Cygni components from all emission lines almost vanished from all lines after Aug 12. As ejecta expand, the density of the nova's atmosphere drops, allowing Lyman $\beta$ photons to reach farther into the ejecta's outermost regions, where neutral oxygen is present. As a result of this Lyman $\beta$ pumping, the permitted emission strength of the 8446 {\AA} increases \citep{2001MNRAS.320..103S}. In the spectrum of RS oph, the strength of \ion{O}{i} 8446 {\AA} was always greater than that at 7774 {\AA}, indicating that the Lyman $\beta$ fluorescence pumping could be the dominant excitation mechanism.

The nebular phase begins when the forbidden lines from highly ionised species in the spectrum become more pronounced. The emission lines of [\ion{O}{iii}] 4363, 4959, and 5007 {\AA} first appeared on August 17.8, which marks the onset of the nebular phase. These emission lines and other forbidden emission lines become more prominent compared to the continuum as the nova ejecta evolve in time. The emission lines of \ion{N}{iii} 4641 {\AA}, and \ion{He}{ii} 4686 {\AA} were already prominent in the spectrum on August 21.91, and the strength of both lines increased after that, while the \ion{Fe}{ii} emission features started becoming weak. On Aug 30.87, 21.94 days after outburst, the [\ion{N}{ii}] 5755 {\AA} line showed an intense broad emission component, but the [\ion{O}{iii}] lines showed just a narrow component, indicating that the electron density of the ejecta might be low enough to display the [\ion{N}{ii}] 5755 {\AA} line but still too high for the [\ion{O}{iii}] lines \citep{Iijima09}.

\subsection{Spectral line profiles}
\begin{figure*}
\centering
\includegraphics[scale=0.63]{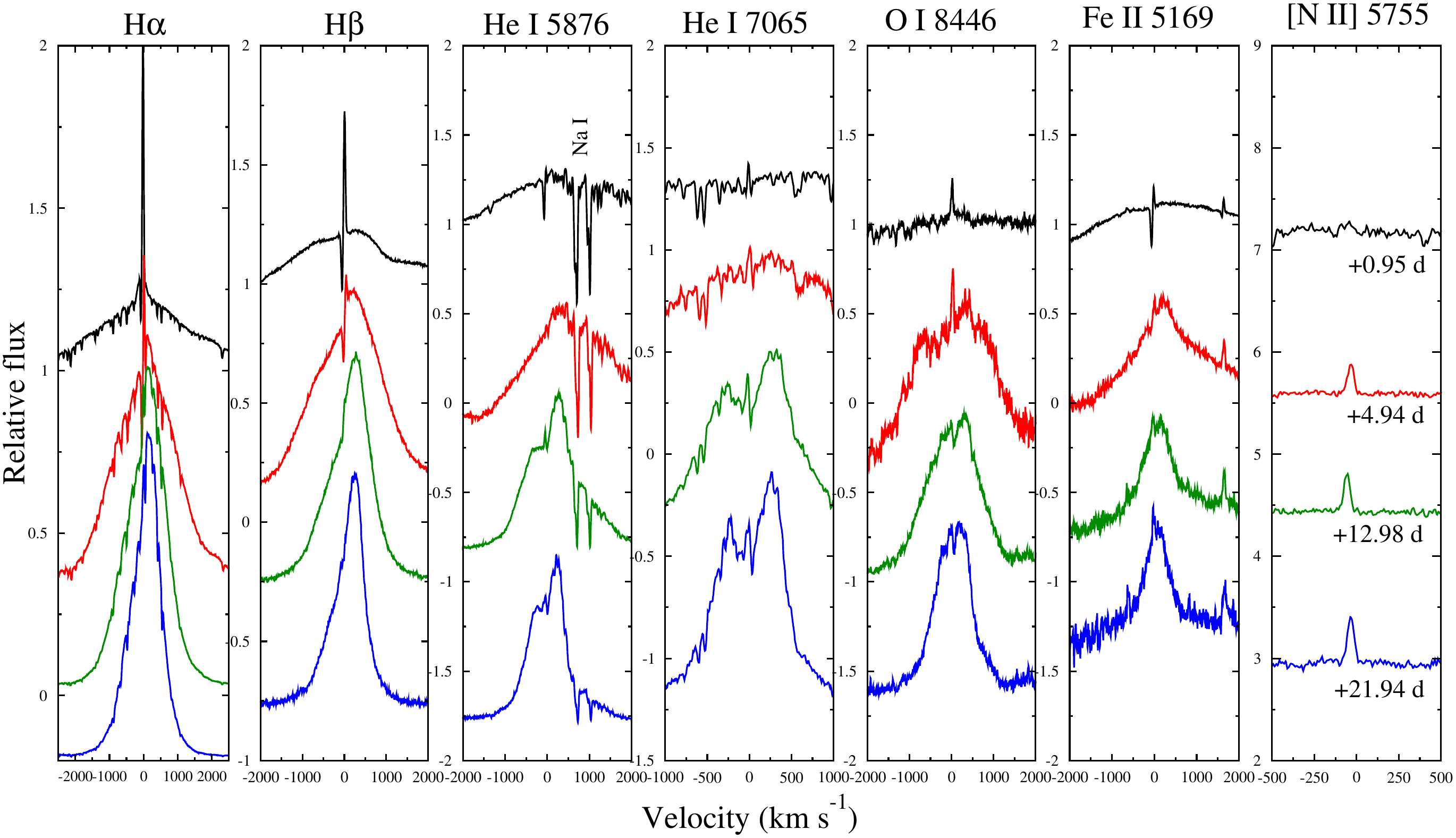}
\caption{The evolution of emission line profiles of \ion{H}{$\alpha$}, \ion{H}{$\beta$}, \ion{He}{i} 5876, \ion{He}{i} 7065, \ion{O}{i} 8446, [\ion{N}{ii}] 5755, and \ion{Fe}{ii} (42) 5169 lines from the first day of the 2021 outburst. The effects of interstellar extinctions are corrected by assuming E(B-V)=0.73. The velocity (km s$^{-1}$) is plotted in abscissa, and the relative flux plus some offset is plotted in ordinate. \label{fig:lines}}
\end{figure*}

The evolution of \ion{H}{$\alpha$}, \ion{H}{$\beta$}, \ion{He}{i} 5876 {\AA}, \ion{He}{i} 7065 {\AA}, \ion{O}{i} 8446 {\AA}, [\ion{N}{ii}] 5755 {\AA}, and \ion{Fe}{ii} (42) 5169 {\AA} line profiles is presented in Figure~\ref{fig:lines}. The spectra of a day after an outburst showed sharp, narrow P Cygni emission features superimposed over the broad emission features. After an outburst, the hydrogen Balmer lines, \ion{Fe}{ii} lines and \ion{He}{i} 5876 {\AA} showed a complex profile composed of a very broad component along with a weak emission component accompanied by a blue-shifted P Cygni absorption component. These narrow P Cygni absorption components originates from the circumstellar medium of the secondary red giant \citep{2004PASJIkeda}. The narrow and broad components were prominent in these lines. \ion{Na}{i} D1 and D2 also had two pairs of narrow absorption components in the spectra. These components are part of the circumstellar envelope generated by the mass outflow from a red giant companion \citep{Iijima09}. The emission line of [\ion{N}{ii}] 5755 {\AA} showed only a narrow component. The evolution of line profiles during the 2021 outburst shows similarity with the previous outburst in 2006 \citep{Iijima09}.

The widths of all emissions lines were broad in the initial days after an outburst, which indicated that the material was ejected from the WD at high velocities, and then they gradually narrowed. This behaviour is usually observed in nova spectra and is probably caused by the velocity distribution in geometrically extended ejecta at the time of the outburst. The gas that moves at the highest velocity travels far more distance in a given time. The ejecta density decreases with the evolution of nova, which causes a decline in the emissivity at any velocity, resulting in a narrowing of the emission lines \citep{1974ApJSStarrfield}. The ionisation of the expanding ejecta increases during this line narrowing stage because the hot central ionising source continues to illuminate and ionise the shell. These characteristics, as well as their evolution, may imply that the ejection occurred as a result of an explosion triggered by TNR rather than an episodic mass transfer event \citep{1996Shore}. This smooth decline in the widths with time is also consistent with the previous outburst in 2006 \citep{2018Mondal}. This behaviour was also observed in the UV during the 1985 outburst \citep{1996Shore}.

\subsection{Propagation of shock wave}\label{sect:shock}
\begin{table}
\centering
\caption[]{Observed widths of the \ion{H}{$\alpha$} and \ion{H}{$\beta$} lines of RS Oph during 2021 outburst.\label{tab:width}}
\setlength{\tabcolsep}{5pt}
\small
\begin{threeparttable}
\centering
\begin{tabular}{ccc}
\hline\noalign{\smallskip}
Days after outburst & FWHM \ion{H}{$\alpha$} (km s$^{-1}$) & FWHM \ion{H}{$\beta$} (km s$^{-1}$) \\
\hline
0.95 & 3878.67 $\pm$ 62.27 & 2931.90 $\pm$ 54.14\\
1.06 & 3753.88 $\pm$ 61.26 & 2951.03 $\pm$ 54.32\\
2.94 & 3516.18 $\pm$ 59.29 & 2695.54 $\pm$ 51.91\\
3.91 & 3087.85 $\pm$ 55.56 & 2660.98 $\pm$ 51.58\\
4.94 & 2556.22 $\pm$ 50.55 & 2185.19 $\pm$ 46.74\\
7.90 & 1931.34 $\pm$ 43.94 & 1588.44 $\pm$ 39.85\\
8.89 & 1788.26 $\pm$ 42.28 & 1483.53 $\pm$ 38.51\\
9.89 & 1677.18 $\pm$ 40.95 & 1448.98 $\pm$ 38.06\\
10.92 & 1575.69 $\pm$ 39.69 & 1414.42 $\pm$ 37.60\\
11.96 & 1480.61 $\pm$ 38.47 & 1278.65 $\pm$ 35.75\\
12.98 & 1416.16 $\pm$ 37.63 & 1226.20 $\pm$ 35.01\\
13.98 & 1382.33 $\pm$ 37.17 & 1174.98 $\pm$ 34.27\\
14.93 & 1306.91 $\pm$ 36.15 & 1105.24 $\pm$ 33.24\\
16.89 & 1199.48 $\pm$ 34.63 & 1021.93 $\pm$ 31.96\\
17.88 & 1163.83 $\pm$ 34.11 & 997.87 $\pm$ 31.58\\
19.87 & 1104.86 $\pm$ 33.23 & 890.49 $\pm$ 29.84\\
20.87 & 1065.55 $\pm$ 32.64 & 913.32 $\pm$ 30.22\\
21.94 & 1036.75 $\pm$ 32.19 & 867.66 $\pm$ 29.45\\
\hline
\end{tabular}
\end{threeparttable}
\end{table}
RS Oph belongs to the class of recurrent novae where the secondary star is an evolved red giant and the primary WD orbits within its wind \citep{1996Shore}. When the ejected material moving outward with a very high velocity is obstructed by the surrounding winds of the red giant companion, it generates a shock wave. This shock wave propagates into the red giant wind. The presence of shock in RS Oph was first detected during the 1985 outburst in the form of bright radio synchrotron emissions \citep{1986ApJ...305L..71H}. \cite{2006Sokoloski} also reported a shock in the RS Oph system in the form of a hot X-ray emitting plasma during the 2006 outburst. \cite{2006DasRSOph} reported the first detection of an infrared shock wave in a recurrent nova.

\begin{figure}
\centering
\includegraphics[scale=0.34]{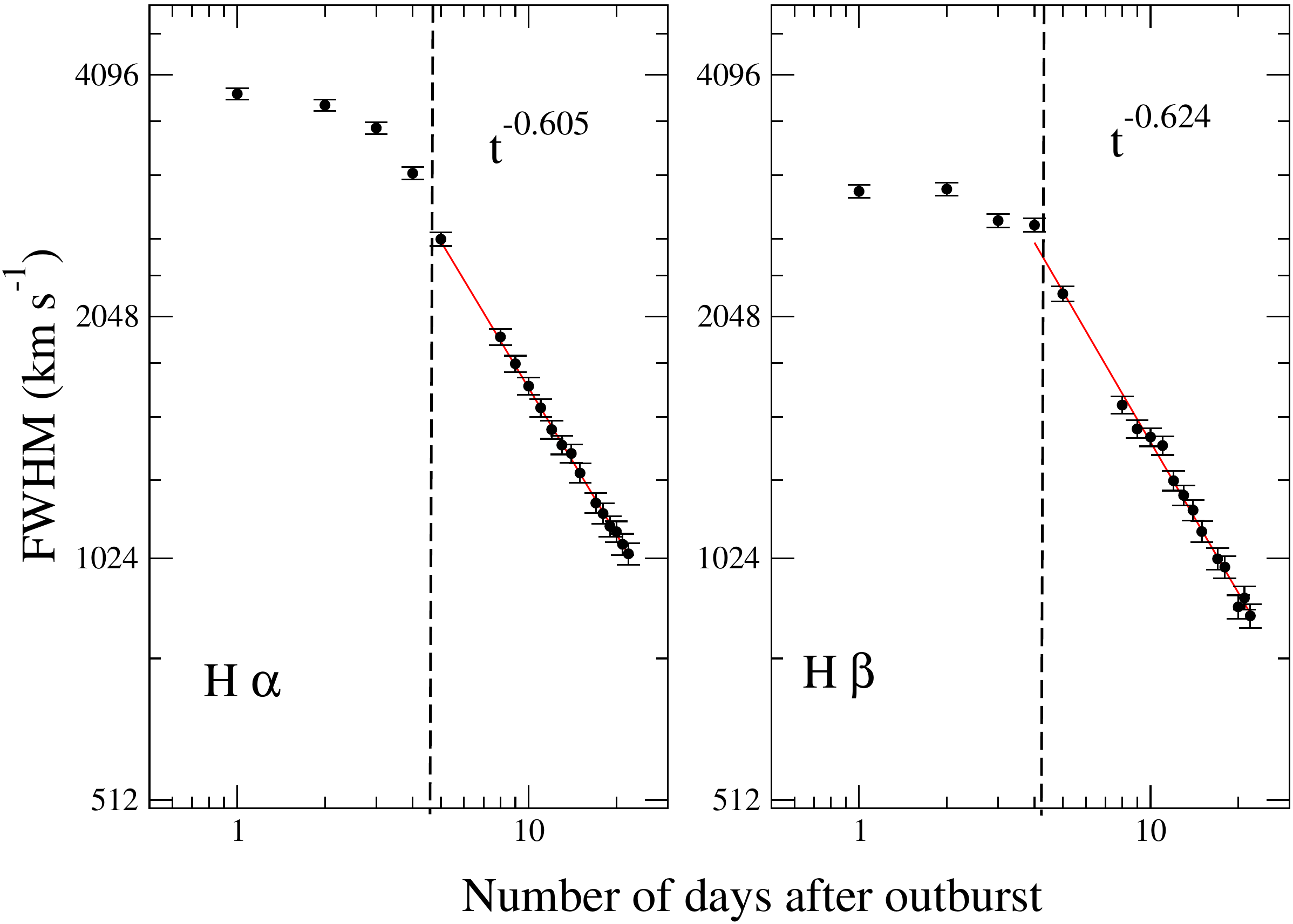}
\caption{The evolution of the line widths for the \ion{H}{$\alpha$} and \ion{H}{$\beta$} lines is shown. The region to the left of the dashed line shows that the ejecta was in free expansion phase for almost 4 days. This phase is followed by a decelerative phase which is best described by a power law. Data points are plotted with $1 \sigma$ error bars. Further details can be found in the text in Section~\ref{sect:shock}. \label{fig:fwhm}}
\end{figure}

We chose two prominent lines, e.g., \ion{H}{$\alpha$} and \ion{H}{$\beta$} emission lines to study the behavior of shock velocity. We measured their line widths by fitting Gaussian using the \textit{splot} task of \textsc{iraf}. We chose these two lines because these two \ion{H}{i} lines are very strong and are not blended with other lines, which ensures the reliable estimation of the line widths. $1\sigma$ error is taken into account when calculating the line widths. The estimated line widths are present in Table~\ref{tab:width}. In this study, the instrumental broadening has not been decoupled from the observed line profiles when calculating their widths, since the instrumental resolution is very high ($R > 9000$), and thus it does not have a significant impact on the measurements. It should be noted that for the early phase of evolution, the measured widths of \ion{H}{$\alpha$} and \ion{H}{$\beta$} might be more uncertain than those presented in Table~\ref{tab:width}. Due to the presence of P Cygni absorption components in the emission lines and the shape of the continuum resulting from \ion{Fe}{ii} multiplet emission in the early phase ($t < 5$ days) , it is difficult to determine the line widths. With the evolution of the nova ejecta in the later phases ($t > 5$days), the ejecta density decreases, the continuum structure decreases, emission lines become more prominent, and these uncertainties with Gaussian fitting become less significant. It should be mentioned here that our primary aim is to only check the variation of line widths with time. The temporal evolution of line widths with time is shown in Figure~\ref{fig:fwhm}. We used a power-law ($t^{-\alpha}$) to fit the data and estimated $\alpha = 0.605$, and 0.624 for \ion{H}{$\alpha$} and \ion{H}{$\beta$}, respectively. The values of $\alpha$ found in this study match well with the previously published values of $\alpha = 0.6$ \citep{2006BodeRSOphUV} and $\alpha = 0.64$ \citep{2006DasRSOph} during the 2006 outburst.

The theoretical model of shock wave propagation in RS Oph, by assuming a simple spherical geometry, has been discussed by \cite{1985BodeKahn}. In their theoretical model, the propagation of shock waves is divided into 3 phases. Phase 1 is the \enquote{ejecta dominated phase} in which a shock at constant velocity is produced as the ejecta expand freely into the winds of the red giant companion. A rapid deacceleration in shock velocity happens during the next two phases. Phase 2 is called \enquote{adiabetic phase}, in which the shock wave is driven into the winds of the red giant, and the shock velocity is expected to behave as $v \propto t^{-1/3}$. There is negligible cooling of the hot shocked material due to radiation losses. In phase 3, the shock material is cooled by the radiation, and shock velocity is expected to behave as  $v \propto t^{-1/2}$ \citep{1985BodeKahn, 2006DasRSOph}. This behaviour is very prominent in Figure~\ref{fig:fwhm} which shows that in the first 4 days, the ejected material was in the free expansion phase, and then moved to the radiative phase after a very brief or undetected adiabatic phase. Similar behaviour was also detected by \cite{2006DasRSOph} in the previous outburst of RS Oph in 2006 using IR observations. The \textit{Swift} observations with X-Ray Telescope (XRT) instruments showed that phase 1 terminated after 6 days and eventually the shock velocity declines monotonically with time and the remnant rapidly evolved to show behavior characteristic of phase 3 during 2006 outburst (see Figure 4 in \cite{2006BodeRSOphUV}).

\subsection{Photoionization model analysis}\label{sect:cloudy}
During an outburst, novae primarily go through two phases according to their spectral evolution. The early phase of expansion is the optically thick phase, which is followed by an optically thin phase. The different conditions present within the ejecta during these two phases require distinct methods to model the observed spectrum. During the optically thick phase, the spectrum is dominated by P Cygni profiles (see Figure~\ref{fig:spectralevolution}). The spectrum during this phase contains information about the temperature and density distribution in the geometrically extended ejecta, which can be studied accurately by the line-blanketed, expanding, spherical, non-LTE model atmosphere code \citep{1992ApJHauschildt, 1995ApJHauschildt, 1997ApJHauschildt, 2001MNRAS.320..103S}. As the nova evolves into an optically thin phase, the ejecta density decreases, continuum emission becomes less important, and the spectrum shows the prominent collisionally excited nebular lines \citep{1998Gehrz}. At this stage, the emission lines in the spectrum become more prominent, and the spectrum during this optically thin phase can be modelled with a photoionization code that incorporates the major heating and cooling mechanisms as well as a large set of predicted emission lines \citep{1998Gehrz,2001MNRAS.320..103S}. Hence, we used spectra after the pre-nebular phase for photoionization modelling.

We used the photoionization code \textsc{cloudy}\footnote{\url{https://www.nublado.org}}, v.17.02 \citep{2017RMxAA..53..385F} to model the observed spectra of RS Oph (2021). \textsc{cloudy} simulates the physical circumstances of non-equilibrium gas clouds subjected to an external radiation source using detailed microphysics. It simultaneously solves the thermal, statistical, and chemical equilibrium equations for a model emission nebula. For a given set of input parameters, \textsc{cloudy} estimates both the intensities and column densities of a very large number ($\sim 10^4$) of spectral lines from non-LTE (Local Thermodynamic Equilibrium), spanning the whole electromagnetic range. The output spectrum could be compared with the observed spectrum to estimate the physical and chemical conditions of the shell. More details about \textsc{cloudy} are present in \cite{2013RMxAA..49..137F}, and references therein. Numerous studies have used \textsc{cloudy} to model other novae systems, e.g., Nova V693 Coronae Austrinae 1981 \citep{1997MNRAS.290...87V}, PW Vulpeculae 1984 \citep{1997MNRAS.290...75S}, Nova LMC 1991 \citep{2001MNRAS.320..103S}, QV Vul \citep{2002ApJ...577..940S}, V382 Vel \citep{2003AJ....125.1507S}, V1974 Cyg \citep{2005ApJ...624..914V}, RS Oph 2006 \citep{2015DasMondal, 2018Mondal}, V838 Her \& V4160 Sgr \citep{2007ApJ...657..453S}, V1065 Cen \citep{2010AJ....140.1347H}, T Pyx \citep{2019A&A...622A.126P}, V745 Sco, and symbiotic star BT Mon \citep{2020MNRASMondal}, V1280 Sco \citep{Pandey_2022} etc.

We assumed central ionising radiation to have a blackbody shape of high temperature ($T_{BB} > 10^4$ K) and luminosity $L$ (erg s$^{-1}$), surrounded by a spherically symmetric expanding shell. The central ionising source provides heat input through photoionization for the surrounding ejecta. The recombination and collisional line cooling balance the heating of the ejected shell. The TNR model suggests that as the nova expands after an outburst, only a portion of the ejected envelope achieves velocities greater than the escape velocity. The remaining ejected material, whose velocity is less than the ejecta escape velocity, returns to the quasistatic equilibrium and forms an envelope around the WD. The matter in the envelope is burned via nuclear reactions, and the radius of the stellar photosphere decreases with constant luminosity. As a result, depending on the mass of WD, the temperature of the nova increases up to several hundred thousand kelvins, and the peak of spectral energy distribution shifts from the visible to the ultraviolet to the X-ray spectrum \citep{2008Krautter}. This marks the beginning of the SSS phase in nova, where nova starts emitting strongly in the soft X-ray domain. The RS Oph 2021 enters the SSS phase on Sep 4 \citep{2021ATelPage}. \citet{2021ATelorio2} used Chandra High Energy X-ray Gratings (HETG) and ACIS camera and XMM-Newton Reflection Grating Spectrographs (RGS) to monitor the nova around 20 and 23 days after the optical maximum and found no bright super soft component in their observations. In this study, we chose spectra before RS Oph entered in the SSS phase for photoionization modelling. It should be emphasised that the blackbody distribution is used for modelling since \textsc{cloudy}, to the best of our knowledge, lacks the needed high temperatures for realistic WD atmospheres. Our primary aim is to estimate the physical and chemical parameters associated with the system during the 2021 outburst of RS Oph. We are not concerned with the detailed spectral energy distribution. The observed spectrum and the modeled spectrum were matched using fluxes relative to the \ion{H}{$\beta$} emission line.

As discussed in Section~\ref{sect:shock}, the secondary star in the RS Oph system is an evolved red giant, and the primary WD orbits within its wind. Such systems are known as symbiotic like recurrent novae \citep{2012BASIShore}. The variation of velocity of the ejected material with time after outburst can be expressed using the following relation (see Figure~\ref{fig:fwhm}),
\begin{equation}\label{eq:velocity1}
V = 
\begin{cases}
  V_0 &  (0 < t < t_0)\\
  V_0 (t/t_0)^{-\alpha} & (t > t_0)
\end{cases}
\end{equation}
where, $V_0$ is the initial expansion velocity of the ejecta and $t_0 \sim 3.91$ is the time when the shock velocity starts to decline. The value of $\alpha$ was estimated around $\sim -0.6$ (see Figure~\ref{fig:fwhm}). The inner and outer radii of the model could be estimated using the relation,
\begin{equation}\label{eq:velocity2}
R = \int_0^t   V(t)dt.
\end{equation}
Using Equation~\ref{eq:velocity1} and the value of $\alpha$, Equation~\ref{eq:velocity2} becomes,
\begin{equation}\label{eq:velocity3}
R = V_0t_0 + \frac{V_0t_0}{0.4}\Big[\Big(\frac{t}{t_0}\Big)^{0.4} - 1 \Big]
\end{equation}
\cite{Mikolajewska2021} found that the expansion velocity of the ejecta using the \ion{H}{$\alpha$} and \ion{H}{$\beta$} P Cyg profiles to be $\sim -4200$ km s$^{-1}$ and $\sim -4700$ km s$^{-1}$ from the absorption center and the blue absorption edge, respectively. \cite{Munari2021} reported that the P Cyg absorption edges are blueshifted by $\sim-3700$ and $\sim-2700$ km s$^{-1}$. For the present study, we used $V_{min}$ = 2700 km s$^{-1}$ and $V_{max}$ = 4700 km s$^{-1}$ and calculated inner radius ($R_{min}$) and outer radius ($R_{max}$) for the ejecta at each epoch using the Equation~\ref{eq:velocity3}.

In our models, the ejecta density was set by the total hydrogen number density, $n(\text{H})$ [cm$^{-3}$] parameter, given by,
\begin{equation}
n(\text{H}) = n(\text{H}^{0}) + n(\text{H}^{+}) + 2n(\text{H}_{2}) + \sum_{\text{other}} n(\text{H}_{\text{other}})~ \text{cm}^{-3},
\end{equation}
where, $n(\text{H}^{0})$, $n(\text{H}^{+})$, $ 2n(\text{H}_{2})$, and $n(\text{H}_{\text{other}})$ represent hydrogen in neutral, ionised, molecular, and all other hydrogen bearing molecules, respectively. We used a radius dependent power-law hydrogen density distribution, $\rho \propto r^{\alpha}$, \cite{1976MNRAS.175..305B} given by,
\begin{equation}\label{eq:nh}
  n(r)=n(r_{\text{in}})(r/r_{\text{in}})^{\alpha}
\end{equation}
where $n(r)$ and $n(r_{\text{in}})$ represent the density of the ejecta at distance $r$ and the inner radius $(r_{\text{in}})$, respectively.
For the present study, we chose the power-law index $\alpha = -3$. This assumption provides a constant mass per unit volume throughout the ejecta. \textsc{cloudy} also uses a radial dependant filling factor which is given by,
\begin{equation}\label{eq:filling}
f(r)=f(r_{\text{in}})(r/r_{\text{in}})^{\beta} 
\end{equation}
where $\beta$ is the exponent of the power-law. The filling factor describes the volume percent occupied by gas. \citet{1989Anupama} examined the optical spectra of RS oph during 1985 outburst and estimated the mass of ejected envelope based on a  filling factor value of around 0.01, and also argued that the filling factor had a smaller value during initial days and increased with time. In our model, we assumed the value of filling factor to be 0.01 and $\beta$ to be 0. The elemental abundances are set by the \textit{abundance} parameter. 

We used $T_{BB}$, $L$, $n(\text{H})$, $f(r)$, and \textit{abundances} of elements as a free input model parameters. We used a broad sample space to vary the values of all input parameters simultaneously in smaller increments to compute the set of synthetic spectra. We varied $T_{BB}$ in the range of $10^{4}-10^{6}$ K, $L$ in the range of $10^{36}-10^{38}$ erg s$^{-1}$, $n(\text{H})$ in the range of $10^8 - 10^{12}$ cm$^{-3}$, simultaneously with the elemental abundances. The upper and lower limits of the free parameters were chosen based on the results obtained in previous studies. We considered those elements only whose emission lines are present in the observed spectra (helium, nitrogen, oxygen, and iron) while all other elements were kept at their solar abundances \citep{2010Ap&SS.328..179G}, and these abundances did not show any significant changes in the model generated spectra. We chose the final model after many iterations of multiple test models across all epochs. We examined all the models visually first and discarded those models which did not match with the observed spectra at all. Finally, we determined the goodness of fit by calculating $\chi^2$, and $\chi^2_{red}$ of the model, given by the following relation,
\begin{equation}
\chi^2 =\sum_{i=1}^n\frac{(M_i-O_i)^2}{\sigma_i^2}
\end{equation}
where $O_i$, $M_i$, $n$, and $\sigma_i$ represent observed line flux ratios, modelled line flux ratios, number of emission lines used in the model, and error in the observed line flux ratios, respectively. In general, $\sigma$ lies in the range between 10\% and 30\%, depending on how strong the line is relative to the continuum, as well as the possibility of blending with the other lines in the spectrum \citep{2010AJ....140.1347H}. The fluxes of each line were measured interactively by fitting Gaussian in the \textit{splot} task of the \textit{onedspec} package in \textsc{iraf}. The reduced $\chi^{2}$ is given by,
\begin{equation}
\chi^{2}_{\text{red}} = \frac{\chi^2}{\nu}
\end{equation}
where $\nu$ represents the number of degrees of freedom (DOF), which is determined by the difference between the number of observed emission lines ($n$) and the number of free parameters ($n_p$), i.e. $\nu$ = $n - n_p$. For an appropriate fit, the values of $\chi^2 \sim \nu$ and $\chi^2_{red}$ should be low, usually between 1 and 2. We simulated the best-fit model by adjusting one parameter at a time while holding the other free parameters fixed at their best-fit model values, until the value of $\chi^2_{red}$ increased to 2. This method gives an approximate uncertainty for the free parameters within $3\sigma$ \citep{2001MNRAS.320..103S, 2007ApJ...657..453S, Pandey_2022}.

It should be noted that the initial attempts to fit the observed spectra reproduced the majority of prominent emission lines but failed to generate the emission lines of high ionisation potential. As a result of the high density of the ejecta, a radiation-bounded shell was formed with a radius of hydrogen recombination slightly larger than the inner radius of the ejecta. Thus, the high ionisation zones in the model shell were small and did not produce sufficient flux. An additional low density component was added to the previous model in order to take into account these lines. As a result of the lower density, the ionising photons penetrate further into the shell, resulting in a more ionised and hotter shell \citep{2003AJ....125.1507S}. Both components had the same physical parameters, except for the hydrogen density and covering factor of the ejecta, which were allowed to vary independently. The covering factor is a fraction of $4\pi$ sr covered by the gas, and it scales with the \textsc{cloudy} line luminosities. We set the covering factors for both components in such a way that their sum is less than or equal to 1. After multiplying the spectrum of the high-density and low-density components by their respective covering factors, the resultant spectrum was generated by adding the spectra from both components.

\begin{figure*}
\centering
\includegraphics[scale=0.65]{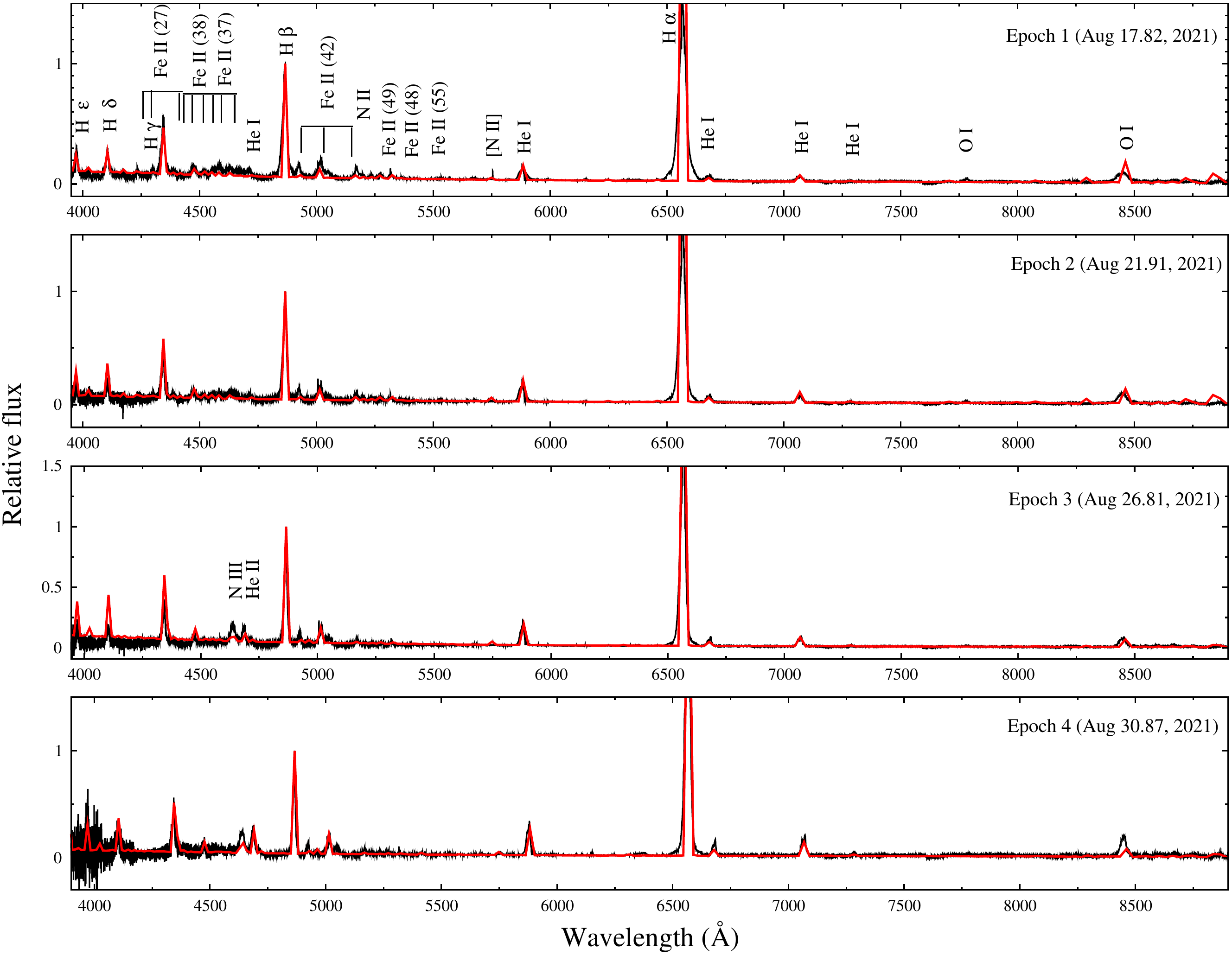}
\caption{Best-fit \textsc{cloudy} model generated spectra (red solid line) plotted over the observed spectra for all epochs. The spectra were normalised to H$\beta$. The observed spectra for all epochs presented here are corrected for reddening by assuming E(B-V)=0.73. \label{fig:cloudymodel}}
\end{figure*}

Table~\ref{tab:parameters} shows the best-fit \textsc{cloudy} model parameters across all epochs. The relative flux of the observed emission lines, the best-fit \textsc{cloudy} model generated line ratios, and their respective $\chi^2$ are are shown in Table~\ref{tab:chisquare}. While calculating the $\chi^2$ values, we have taken into account only those lines that appear in both the \textsc{cloudy} model-generated and observed spectra. The observed emission lines are de-reddened using E(B-V) = 0.73 \citep{1987snijders} and matched with the model spectra. For comparison, the \textsc{cloudy} model generated best-fit spectra (red solid lines) are shown overlaid on observed optical spectra (black lines) for epochs 1, 2, 3 and 4 are presented in Figure~\ref{fig:cloudymodel}. In the figures, the prominent emission lines are marked. The observed line fluxes have been determined interactively using the \textit{splot} task of \textit{onedpsec} package within \textsc{iraf} by fitting each line with a gaussian. The \textsc{cloudy} modeled and observed flux ratios have been calculated relative to \ion{H}{$\beta$}. The $\chi^2_{red}$ values for each epochs are given in Table~\ref{tab:parameters}. The low $\chi^2_{red}$ values suggest the fits are satisfactory. We estimated the values of different parameters by using the best-fit model. There are just a few factors that may significantly affect the spectral characteristics, and even modest adjustments to those input parameters cause the features to change significantly. We discuss these below in detail.

\begin{table*}
\caption{Best-fit \textsc{cloudy} model parameters during the 2021 outburst.\label{tab:parameters}}
\setlength{\tabcolsep}{17pt}
\small
\begin{threeparttable}
\centering
\begin{tabular}{l c c c c c c c ccccc}
\hline
\hline
 Parameters                 &	 Epoch 1   &	Epoch 2  	& Epoch 3 		&  Epoch 4	 \\
                                   &	  Day 09   &	Day 13  	&   Day 18 		&     Day 22	 \\ 

\hline
Temperature of central ionsing WD  ($\times 10^4$K)       & 4.16 $\pm$ 0.25  &  4.89	$\pm$ 0.20	&  6.45	$\pm$ 0.09 & 6.60 $\pm$ 0.08 \\
Luminosity of central ionising WD ($\times 10^{37}$erg s$^{-1}$) & 1.00     & 1.00  	& 1.00		& 1.00  		     \\
Density of dense region of ejecta ($\times 10^{10}$cm$^{-3})$  &   1.47 $\pm$ 0.06 &  0.79 $\pm$ 0.35 & 0.50 $\pm$ 0.04 & 0.35 $\pm$ 0.05 \\
Covering factor (dense region)  & 0.85 & 0.75 & 0.63 & 0.61 \\
Density of hot region of ejecta ($\times 10^8$cm$^{-3})$&  5.62 $\pm$ 0.13  &  5.01  $\pm$ 0.23  & 2.51 $\pm$ 0.25 & 1.77 $\pm$ 0.07  \\
Covering factor (hot region) & 0.15  & 0.25	& 0.37 & 0.39 \\
$\alpha^{a}$	          &   -3     &   -3		& -3	  	&  -3	 \\
R$_{in}^{b}$  ($\times 10^{14}$ cm )	 & 1.81 &  2.29	& 2.81 	& 3.16  \\
R$_{out}^{b}$ ( $\times 10^{14}$cm)          &  3.16  & 3.98 	& 4.89	& 5.49 \\
Filling factor$^a$	              &	  0.01	 &  0.01  	& 0.01   		& 0.01   \\
$\beta^{a}$		          &	0.0	     &  0.0  	& 0.0  		& 0.0  \\
He$^{c}$ &  1.4 $\pm$ 0.30  &  1.4 $\pm$ 0.30  & 1.5 $\pm$ 0.25 &  1.9 $\pm$ 0.20  \\
N$^{c}$        &   70.0$^{+20}_{-15}$  & 70.0$^{+20}_{-15}$  & 90.0 $\pm$ 15 &  95 $\pm$ 15 \\
O$^{c}$      &  0.6 $\pm$ 0.02 &  0.7 $\pm$ 0.03	& 1.9 $\pm$ 0.15 & 2.6 $\pm$ 0.3  \\
Fe$^{c}$		&  1.9 $\pm 0.40$  &  1.9 $\pm 0.40$ & 1.0 $\pm$ 0.20 & 1.0 $\pm$ 0.25 \\
Number of observed lines (n)  &    26    &   26     & 27     &  25	\\
Number of free parameters (n$_{p}$) &  10   &   10     &  10        & 	10	 \\
Degrees of freedom ($\nu$)		  &  16  &	16     &	17    &   15	\\
Ejected mass ($\times 10^{-6} M_{\odot}$) & 3.83 & 3.64 & 3.62 & 3.54 \\
Total $\chi^{2}$ & 29.35 & 30.87 & 27.38 & 24.95	\\
$\chi^{2}_{red}$ & 1.83 & 1.92 & 1.61 & 1.66 \\
\hline
\end{tabular}
\begin{tablenotes}
\item (a) This was not a free parameter in the model.
\item (b) Inner and outer radii were calculated using Equation~\ref{eq:velocity3}. This was not a free parameter in the model.
\item (c) The log abundance by number relative to hydrogen, relative to solar values. The logs of solar abundances relative to hydrogen are: He=-1.07, N=-4.17, O=-3.31, and Fe=-7.44. All other elements that are not listed in the table were set to their solar values.
\end{tablenotes}
\end{threeparttable}
\end{table*}

\begin{landscape}
\begin{table}
\setlength{\tabcolsep}{6pt}
\begin{center}
\caption{Observed and best-fit \textsc{cloudy} model line fluxes.\label{tab:chisquare}}
\begin{tabular}{l l c c c c c c c c c c c c}
\hline\noalign{\smallskip}
 &	      &          &	Epoch 1 &	         &	 	&  Epoch 2 &            &       &	Epoch 3  &	 	      &       &  Epoch 4  &  \\
$\lambda$  (\AA)  & Line ID  & Observed Flux     &  Model Flux    & $\chi^{2}$ & Observed Flux &  Model Flux     & $\chi^{2}$ & Observed Flux  &  Model Flux     & $\chi^{2}$ & Observed Flux  & Model Flux     & $\chi^{2}$\\
\hline\noalign{\smallskip}
3889 & \ion{H}{$\zeta$}, \ion{He}{i} & ... & ... & ... & 0.33E+00 & 0.11E+00 & 0.52E+00 & 0.14E+00 &  0.27E+00 &  0.27E+00 & ... & ... & ... \\
3970    & \ion{H}{$\epsilon$} &  0.20E+00 &  0.15E+00 &  0.30E-01 & 0.23E+00 & 0.13E+00 &  0.11E+00 & 0.19E+00 & 0.19E+00 & 0.53E-03 & ... & ... & ...  \\
4026    & \ion{He}{i}, \ion{He}{ii}  &  0.20E-01 & 0.19E+00 & 0.34E+00 & 0.86E-01 & 0.49E-01 & 0.15E-01 & ... & ... & ... & ... & ... & ...\\
4101    & \ion{H}{$\delta$}   & 0.21E+00 &  0.17E+00 &  0.12E-01 & 0.14E+00 & 0.18E+00 & 0.17E-01 & 0.24E+00 & 0.26E+00 & 0.30E-02 & 0.79E+00 & 0.26E+00 & 0.45E+01\\
4180 & \ion{Fe}{ii}, [\ion{Fe}{ii}] & 0.67E-01 & 0.40E-01 & 0.85E-02 & ... & ... & ... & ... & ... & ... & ... & ... & ...\\
4233 & \ion{Fe}{ii} & 0.45E-01 & 0.44E-01 & 0.71E-05 & ... & ... & ... & 0.16E+00 & 0.15E-01 & 0.37E+00 & ... & ... & ...  \\
4340 & \ion{H}{$\gamma$} & 0.44E+00 & 0.54E+00 & 0.10E+00 & 0.44E+00 &  0.56E+00 & 0.16E+00 & 0.40E+00 & 0.56E+00 & 0.41E+00 & 0.59E+00 & 0.48E+00 & 0.19E+00 \\
4415 & [\ion{Fe}{ii}] & 0.34E-01 & 0.29E-01 & 0.27E-03 & 0.95E-01 & 0.51E-01 & 0.21E-01 & ... & ... & ... & ... & ... & ...\\
4471 & \ion{He}{i} & 0.13E+00 & 0.70E-01 & 0.47E-01 & 0.15E+00 & 0.98E-01 & 0.32E-01 & 0.14E+00 & 0.83E-01 & 0.58E-01 & 0.21E+00 & 0.75E-01 & 0.32E+00\\
4584 & \ion{Fe}{ii} & 0.11E+00 &  0.77E-01 & 0.14E-01 & 0.80E-01 &  0.31E-01 & 0.26E-01 & 0.71E-01 & 0.29E-01 & 0.27E-01 & 0.11E+00 & 0.16E-01 & 0.14E+00 \\
4640 & \ion{N}{iii} & ...  & ... & ... & ... & ... & ... & 0.23E+00 & 0.97E-01 & 0.29E+00 & 0.32E+00 & 0.10E+00 & 0.77E+00 \\
4686 & \ion{He}{ii} & ... & ... & ... & 0.19E+00 & 0.58E-01 & 0.20E+00 & 0.16E+00 & 0.11E+00 & 0.49E-01 & 0.29E+00 & 0.21E+00 & 0.88E-01\\
4713 & \ion{He}{i} & 0.75E-01 & 0.80E-01 & 0.22E-03 & 0.78E-01 & 0.28E-01 & 0.27E-01 & ... & ... & ... & 0.60E-01 & 0.52E-01 & 0.10E-02\\
4863 & \ion{H}{$\beta$}	& 0.10E+01 & 0.10E+01 & 0.00E+00 & 0.10E+01 & 0.10E+01 & 0.00E+00 & 0.10E+01 & 0.10E+01 & 0.00E+00 & 0.10E+01 & 0.10E+01 & 0.00E+00  \\
4922 & \ion{Fe}{ii}, \ion{He}{i} & 0.77E-01 & 0.59E-01 & 0.37E-02 & 0.11E+00 & 0.99E-01 & 0.13E-02 & 0.10E+00 & 0.32E-01 & 0.90E-01 & 0.13E+00 & 0.44E-01 & 0.13E+00  \\
4959 & [\ion{O}{iii}] &...  & ...& ...& ...	& ...&...	&  0.37E-02 & 0.32E-01 & 0.13E-01 & 0.56E-02 & 0.40E-01 & 0.19E-01\\
5016 & \ion{Fe}{ii}, \ion{He}{i} & 0.21E+00 & 0.18E+00 & 0.97E-02 & 0.18E+00 & 0.98E-01 & 0.76E-01 & 0.19E+00 & 0.87E-01 & 0.17E+00 & 0.91E-01 & 0.14E+00 & 0.43E-01 \\
5169 & \ion{Fe}{ii} & 0.95E-01 & 0.79E-01 & 0.27E-02 & 0.77E-01 & 0.53E-01 & 0.60E-02 & 0.72E-01 & 0.15E+00 & 0.11E+00 & 0.10E+00 & 0.20E-01 & 0.11E+00\\
5235 & \ion{Fe}{ii} & 0.20E-01 & 0.67E-01 & 0.24E-01 & 0.56E-01 & 0.67E-01 & 0.12E-02 & 0.10E-01 & 0.10E-01 & 0.49E-07 & 0.22E-01 & 0.18E-01 & 0.20E-03 \\
5276 & \ion{Fe}{ii} & 0.23E-01 & 0.38E-01 & 0.27E-02 & 0.53E-01 & 0.62E-01 & 0.10E-02 & 0.89E-02 &  0.16E-01 & 0.95E-03 & 0.36E-01 & 0.22E-01 & 0.31E-02 \\
5317 & \ion{Fe}{ii} & 0.33E-01 & 0.40E-01 & 0.49E-03 & 0.45E-01 & 0.50E-01 & 0.22E-03 & 0.25E-01 & 0.19E-01 & 0.50E-03 & 0.43E-01 & 0.17E-01 & 0.10E-01 \\
5412 & \ion{He}{ii} & ...  & ...& ...& ...	& ...&... & 0.12E-01 & 0.93E-02 & 0.15E-03 & 0.48E-01 & 0.34E-01 & 0.29E-02\\
5535 & \ion{N}{ii} & 0.74E-02 & 0.52E-02 & 0.55E-04 & 0.11E-01 & 0.25E-01 & 0.24E-02 & 0.86E-02 &  0.25E-02 & 0.59E-03 & 0.20E-01 & 0.19E-01 & 0.41E-04\\
5755 & [\ion{N}{ii}] & 0.25E-02 & 0.15E-01 & 0.17E-02 & 0.10E-02 & 0.23E-01 & 0.58E-02 & 0.43E-02 & 0.17E-01 & 0.26E-02 & 0.21E-01 & 0.22E-01 & 0.23E-04\\
5876 & \ion{He}{i} &  0.16E+00 & 0.12E+00 & 0.14E-01 & 0.25E+00 & 0.21E+00 & 0.20E-01 & 0.28E+00 &  0.27E+00 & 0.29E-02 & 0.43E+00 & 0.31E+00 & 0.20E+00\\
6563 & \ion{H}{$\alpha$} & 0.24E+01 & 0.42E+01 & 0.26E+02 & 0.28E+01 & 0.45E+01 & 0.29E+02 & 0.32E+01 & 0.45E+01 & 0.24E+02 & 0.60E+01 & 0.49E+01 & 0.16E+02\\ 
6678 & \ion{He}{i} & 0.61E-01 & 0.52E+00 & 0.23E+01 & 0.12E+00 & 0.82E-01 & 0.17E-01 & 0.16E+00 &  0.10E+00 & 0.48E-01 & 0.26E+00 & 0.12E+00 & 0.33E+00\\
7065 & \ion{He}{i} & 0.71E-01 & 0.61E-01 & 0.11E-02 & 0.12E+00 & 0.10E+00 & 0.23E-02 & 0.23E+00 &  0.17E+00 & 0.57E-01 & 0.49E+00 & 0.33E+00 & 0.40E+00\\
7280 & \ion{He}{i} & 0.38E-01 & 0.28E-01 & 0.11E-02 & 0.13E-01 & 0.12E-01 & 0.39E-05 & 0.47E-01 & 0.16E-01 & 0.15E-01 & 0.83E-01 & 0.26E-01 & 0.51E-01\\
7774 & \ion{O}{i} & 0.39E-01 & 0.10E-01 & 0.93E-02 & 0.26E-01 & 0.35E-02 & 0.57E-02 & 0.14E-01 &  0.42E-02 & 0.16E-02 & 0.34E-01 & 0.98E-02 & 0.96E-02\\
8448 & \ion{O}{i} &  0.18E+00 & 0.25E+00 & 0.51E-01 & 0.22E+00 & 0.25E+00 & 0.11E-01 & 0.18E+00 & 0.31E-01 & 0.38E+00 & 0.55E+00 & 0.29E+00 & 0.11E+01\\
\hline
Total &	 & & & 29.35  & & &  30.87  & & & 27.38 & & & 24.95\\
\noalign{\smallskip}\hline
\end{tabular}
\end{center}
\end{table}
\end{landscape}

\subsubsection{Temperature and luminosity of the central ionising source}
The temperature of the central ionising source from the best-model is mentioned in Table~\ref{tab:parameters}. The temperature of the central ionising WD during epoch 1 was determined to be $\sim 4.16 \times 10^4$ K, which increased to $\sim 6.60 \times 10^4$ K during epoch 4, probably due to the collapse of the pseudo-photosphere onto the surface of WD as the nova evolved \citep{1976MNRASBath&Shaviv}. The luminosity of the central ionising WD from the best-fit model was determined to be $\sim 1.0 \times 10^{37}$erg s$^{-1}$ for all epochs. We found that the variation in the luminosity parameter had only a marginal effect on the model-generated spectrum. Novae atmospheres are optically thick and fast-expanding shells with gradually decreasing densities as their radius increases. As a result, very large geometrically extended atmospheres develop, which have large temperature differences between the inner and outer parts of line-forming regions \citep{1992ApJHauschildt}. A change in the relative geometric extension of the atmosphere affects the radiative transfer and temperature structures of the ejecta. And for the $\alpha = -3$ (see Section~\ref{sect:cloudy}), the geometrical extension does not change significantly with the luminosity parameter. The effects are caused by changes in density parameters due to spherical geometry \citep[ch-5]{2008clno.bookBode}. \cite{2018Mondal} also reported a similar value of luminosity of the central WD ($\sim 10^{37}$erg s$^{-1}$) in their photoionization modelling of the 2006 outburst of RS Oph.

\subsubsection{Ejecta density}
The density of both components from the best-fit model is mentioned in Table~\ref{tab:parameters}. The ejecta density for both components decreased from epoch 1 to epoch 4. As the ejecta expanded, the ejecta density decreased, allowing ejecta to be exposed to underlying radiation from the central ionising WD. This could be inferred by the appearance of emission lines of species with high ionisation potential. In the present study, we observed that in order to generate the observed \ion{Fe}{ii} multiplets between 4500 {\AA} and 5000 {\AA}, during epochs 1 and 2, a high ejecta density ($n(H)> 7.00 \times 10^{9}$cm$^{-3}$) was required. Additionally, the emission feature of \ion{O}{i} 8446 {\AA} also appeared from the region with high ejecta density ($n(H) > 2.52 \times 10^{9.0}$cm$^{-3}$). As ejecta expand, more Lyman $\beta$ photons can penetrate the outer regions of the ejecta, which are rich in neutral oxygen. This Lyman $\beta$ pumping increased the strength of \ion{O}{i} emission line at 8446 {\AA}. The strength of \ion{O}{i} 8446 {\AA} further increased with the increase in the ejecta density of the dense component. As the ejecta evolved and the density decreased, the ionisation of ejecta increased and the emission features from \ion{He}{i} 5876 {\AA}, \ion{He}{ii} 4686 {\AA},and \ion{N}{iii} 4640 {\AA} became more prominent. These features were well reproduced from the region of low hydrogen density of the ejecta. The \ion{N}{iii} 4640 {\AA} feature in the spectra of epochs 3 and 4 was very sensitive to the N abundance and the density of the ejecta. We found that this feature appeared for the ejecta density between $1.58 \times 10^{8}$ to $2.81 \times 10^{8}$cm$^{-3}$. After that, the strength of \ion{N}{iii} 4640 {\AA} decreased with the increasing density even with an increase in N abundance. The [\ion{N}{ii}] 5755 {\AA} feature was also very sensitive to the ejecta density of the less dense component. We found that this emission feature appeared for an ejecta density $> 1.00 \times 10^{8}$cm$^{-3}$, and its strength increased with an increase in the density. For the density $>10^{9}$cm$^{-3}$, the strength of [\ion{N}{ii}] 5755 {\AA} line started to decrease. We carefully examined all these variations to set the density of the model.

\subsubsection{Abundance of the ejecta}
The elemental abundances determined from the best-fit model are presented in table~\ref{tab:parameters}. 
The He abundance was determined by the fitting of prominent \ion{He}{i} (4471, 4713, 5876, 6678, 7280 {\AA}), \ion{He}{ii} 4686 {\AA} lines in all epochs. The low-density part of the ejecta was required to generate these He lines. We found the value of He $\sim 1.4$ is required to generate the observed He lines during epoch 1, which further increased to $\sim 1.9$ during epoch 4. The high value of He abundance is consistent with the TNR scenario for RS Oph, which indicates the accretion onto a very massive WD with a mass close to the Chandrashekar limit produces a significant amount of helium \citep{1995MNRAScontini}. To estimate N abundance, we used the emission lines of \ion{N}{ii} 5535 {\AA}, [\ion{N}{ii}] 5755 {\AA} during epochs 1, and 2, \ion{N}{iii} 4640 {\AA} during epochs 3 and 4. From our best-fit model, we found that a high value of N $\sim$ 70-95 was required to generate these lines. The emission line of \ion{N}{iii} 4640 {\AA} was sensitive to the changes in temperature of the central ionising WD, ejecta density of the less dense components, and the N abundance. We varied these parameters simultaneously to generate these lines and found that a relatively low hydrogen density of $\sim 1.58 \times 10^{8} - 2.81 \times 10^{8}$cm$^{-3}$, and the temperature of central ionising WD around $\sim 5.0 \times 10^{4} -  7.07 \times 10^{4}$ K, and the N abundance of 70-95 is required. The \ion{N}{iii} line was not generated in the modelled spectra at higher temperatures $(> 8 \times 10^4 \text{K})$. To determine O abundance, we used the \ion{O}{i} line at 8446 {\AA}. We found that the O abundance is enhanced over its solar value. Fe abundance was estimated by the fitting of \ion{Fe}{ii} multiplets present in the optical spectra. We found an enhanced value for Fe over its solar value. The enhanced Fe abundance value could imply that the nova ejecta are helium-rich and that the mass-losing star is more evolved than the main sequence star \citep{2008clno.bookBode}. Novae ejecta often have enhanced metal abundances, which are commonly observed and predicted by theoretical models as well \citep{1994ApJLivio}. According to multidimensional analyses of TNR, the most plausible explanation for the elemental enhancement in the nova ejecta is the result of adequate mixing of accreted material from the secondary with the underlying WD at the core envelope interface during the TNR \citep[for e.g. see][and references therein]{1998Gehrz,2012BASIJose,2020A&AJose,2020ApJStarrfield}.

In the previous 2006 outburst, \cite{2018Mondal} determined the abundances by optical spectral fitting. The values reported in the present study are in a similar ballpark to the values reported by them. However, we found a very high value of N abundance in the present study. Here, we focused on the \ion{N}{iii} 4640 {\AA}, \ion{N}{ii} 5535 {\AA}, and [\ion{N}{ii}] 5755 {\AA} lines in the optical spectra to determine the N abundance. The model of \cite{2018Mondal} couldn't generate the \ion{N}{iii} 5755 {\AA} line present in the observed spectra during the 2006 outburst. Based on the optical study, the helium abundance of He/H$_{\odot}$ = 0.16 was reported during 1985 outburst \citep{1989Anupama}. From Spitzer and ground-based observations, \cite{2007ApJEvans} reported O/Ne $>$ 0.6 by number, during the 2006 outburst. Study of the time development of \ion{N}{v} 1240 {\AA} in UV spectra from IUE during the 1985 outburst provided evidence that N might be overabundant \citep{1996Shore}. \cite{2017ApJ...847...99M} estimated an enhancement of N/O by a factor of $10 - 20$ with respect to solar and the depletion of C/N to 1/40 solar, which implies an enrichment of N and a depletion of C in CNO cycle burning. A very high value of N/O is also consistent with an outburst on a very massive WD \citep{1994ApJLivio}. In the cases, where outburst occurs on the surface of a very massive WD ($\sim 1.4M_{\odot}$), both carbon and oxygen are converted into nitrogen at the peak of TNR and both carbon and oxygen abundance becomes significantly low \citep{1995MNRAScontini}. \cite{1987snijders} reported $n(C)/n(N)=1.10 \pm 0.17$, $n(C)/n(N)=0.16\pm 0.04$, and a substantial overabundance of nitrogen over helium. They also argued that the material transferred from the red giant could be enriched in nitrogen. Additionally, N was also found to be enhanced in the atmosphere of the red giant companion of the RS Oph system \citep{2008A&APavlenko}. \citet{2009ApJStarrfield} also argued that the observed enhanced nitrogen in the nova ejecta is probably $^{15}\text{N}$, and it was produced during earlier outbursts, incorporated into newly accreted material from the secondary star, and then ejected during the present novae outburst. \cite{2009ApJDrake} studied the early pre-SSS Chandra High Energy Transmission Grating (HETG) spectra of RS Oph from the outburst in the year 2006, and argued that the ejecta might contain super-solar abundances. A more detailed discussion on the elemental abundances of RS Oph using two independent approaches of modelling of spectra taken from Chandra and XMM-Newton is presented by \cite{2009AJNess}. A compilation of abundances studied in novae ejecta is presented in \cite{1998Gehrz}, \citep[ch-6]{2008clno.bookBode}.

\subsubsection{Ejected mass}
The hydrogen mass contained in a model ejecta could be estimated using the following relation from \cite{2001MNRAS.320..103S},
\begin{equation}
    M_{\text{shell}} = \int n(r)f(r) dV(r)
\end{equation}
where, n(r), and f(r) are the hydrogen density (cm$^{-3}$) and filling factor, respectively. Using Equation~\ref{eq:nh}, and \ref{eq:filling},
\begin{equation}
M_{\text{shell}} = n(r_{\text{in}})f(r_{\text{in}}) \int_{r_{\text{in}}} ^{r_{\text{out}}} (r/r_{\text{in}})^{\alpha + \beta} 4\pi r^2 dr
\end{equation}
The density, filling factor, the values of $\alpha$, and $\beta$ of the shell are set by the best-fit model density (Table~\ref{tab:parameters}). For the present study, the \textsc{cloudy} model assumed two components, and the total shell mass could be estimated by multiplying the shell mass from both components with their respective covering factors and then adding them. The \textsc{cloudy} best-fit model for epochs 1, 2, 3, and 4, predicted the ejected hydrogen shell mass of $\sim 3.83 \times 10^{-6} \text{M}_{\odot}$, $3.64 \times 10^{-6} \text{M}_{\odot}$, $3.62 \times 10^{-6} \text{M}_{\odot}$, and $3.54 \times 10^{-6} \text{M}_{\odot}$, respectively. It should be noted that we only estimated the ejected hydrogen shell mass using the best-fit \textsc{cloudy} model paramaters. Our model predicted an overabundance of elements in the ejecta. As a result, the reported numbers may represent the lower limit for ejected mass in RS Oph during the 2021 outburst.

During the previous outburst in 2006, \cite{2015DasMondal} derived the ejected mass of $3.4 \times 10^{-6} M_{\odot}$ and $4.9 \times 10^{-6} M_{\odot}$, for days 31 and 49, respectively. In the study by \cite{2006DasRSOph} in IR, the ejected shell mass was estimated to have a mean value of $\sim 3 \times 10^{-6}M_{\odot}$ during the 2006 outburst. \cite{2009A&Aorlando} derived the mass of ejecta to be $\sim 10^{-6} M_{\odot}$ from Chandra/HETG data during the 2006 outburst. During 1985 outburst, \citet{1989Anupama} estimated the mass of the ejected envelope around $(3.1 \pm 0.6) \times 10^{-6}$ M$_{\odot}$. \cite{2017ApJ...847...99M} showed that the RS Oph system hosts a CO type WD of mass in the range $1.2 - 1.4 M_{\odot}$ and the maximum mass of a CO WD due to stellar evolution cannot exceed $1 - 1.1 M_{\odot}$; hence, the WD is likely to have grown to its present value as a result of accretion. The 3D hydrodynamic simulations of quiescent accretion with the subsequent explosive phase of RS Oph by \cite{2008A&AWalder} showed that the mass of WD in RS Oph is increasing with time. In a study by \cite{2007ApJHachisu}, they estimated the WD mass to be $1.35 \pm 0.01 M_{\odot}$, and the growth rate of WD mass to be at an average rate of $\sim 1 \times 10^{-7} M_{\odot}$ yr$^{-1}$. The low ejected mass could indicate that the WD retains most of its accreted envelope and that the WD continues to grow in mass, which probably explodes as a Type Ia supernova. In a recent study of TNR, \citet{2020ApJStarrfield} studied the evolution of TNR on CO WD by considering a mixing of 25\% CO WD and 75\% solar matter when the TNR is already in progress, and showed that only a portion of accreted mass is expelled during outburst, such that the mass of CO WD mass is increasing.

\subsection{3D morpho-kinematic modeling of the ejecta}\label{sect:3D}
We predicted the morphology, density distribution, and expansion velocity field of the ejecta of RS Oph during the 2021 outburst by modelling the emission line profiles in the spectra. The line profiles from an expanding shell carry information on the correlation of morphology and kinematic characteristics along the line-of-sight. The correlation is depicted in 1D or 2D spectral profiles, from which we can obtain position-velocity (PV) profiles (e.g., \citet{2009ApJRibeiro}). These profiles characteristically provide the position of a point within the ejecta along with its velocity. In this work, we used the \ion{H}{$\alpha$} PV spectral profile to reconstruct the 3D structure of the ejecta. The line profiles from nova ejecta are broad due to high expansion velocity. Hence, the PV profiles generally resolve the PV correlation features along the line-of-sight. The aim is to obtain a 3D model with a specific density distribution and expansion velocity field, such that the model-generated PV profile would fit the observed PV profile. We used the 3D modelling code \textsc{shape} \citep{2011Shape} to obtain the model. In \textsc{shape}, we can construct a 3D multi-component mesh model using basic geometries, e.g., sphere, torus, cone, etc. The basic geometry can be modified using various structural modifiers, e.g., squeeze, spiral, twist, etc. The components are attributed with a density profile and velocity field using density and velocity modifiers. The code computes radiation transfer through the nebula and generates model images and position-velocity diagrams. In the present study, a large number of models with different configurations were computed, and finally, the best-fitting model was obtained. 

First, we inspected the qualitative characteristics of the profiles in RS Oph spectra during the 2021 outburst. The profiles have similarities with that of the RS Oph 2006 outburst. All the profiles show the presence of a blue-shifted and a red-shifted peak around the central maximum. The blue and red peaks are quite prominently visible in a few profiles, e.g., in the \ion{O}{i} profile. However, in some profiles, for example, the \ion{H}{$\alpha$} profile, the peaks are almost fused with the central maximum. The profiles also suggest the presence of two more wings with high blue and redshifts from their shapes near the continuum. 

For the 3D morpho-kinematic modeling, we chose the \ion{H}{$\alpha$} spectral profiles on Aug. 16.83 and 30.87, 2021. This provides an understanding of the 3D morphology and dynamics of the ejecta as well as the evolution of the same during the first few days of the outburst. \citet{2009ApJRibeiro} modelled the ejecta at its expanded phase of RS Oph 2006, with a bipolar structure with density enhancement around the waist. They obtained an inclination of ${39^{+1}_{-9}}^{\circ}$ of the ejecta axis (same as the binary axis). \citet{1996Shore} reported the inclination angle as $30^{\circ}<i<40^{\circ}$. \citet{1994AJDobrzycka} also reported the inclination angle for the central binary to be $30^{\circ}<i<40^{\circ}$. In the previous 2006 outburst, the angle of the binary system was estimated to be $30^{\circ} - 40^{\circ}$ \citep{2006Sokoloski}. Hence, based on the literature, we constructed our models assuming the inclination range of $30^{\circ}<i<40^{\circ}$. The 3D structure of the ejecta in a different outburst of a nova is expected to be similar as the outburst mechanism is expected to be unchanged. The similarities in spectral profiles between the 2006 and the present outburst suggest the same as well. However, here we aim to model the ejecta at an early phase of the outburst, which ought to have a difference in density distribution and shape than that of an expanded phase. Hence, we performed an independent reconstruction of the 3D morphology and density distribution of the ejecta, with no particular reference to the earlier RS Oph models.

We used normalised \ion{H}{$\alpha$} profiles on Aug. 16.83 and 30.87, 2021 to reconstruct our 3D model. Our first attempt was to construct the basic 3D structure of the ejecta with the combination of shapes and components that are commonly observed in novae ejecta in general: elliptical or bipolar shaped filled structures or shells, central disk/toroid, polar caps, etc, (e.g., \citet{2011MNRASRibeiro}). Our aim was to fit the characteristic features observed in the profile so that the final modelled profile would reproduce the observed shape, both intensity (height) and velocity (width), of the profile as closely as possible. Hence, we varied the density profile of the ejecta to match the height of the profile for different velocity values within the profile. We assumed an expansion velocity field $v(r) \propto r$ for the ejecta. We also varied the magnitude of the velocity in order to match the width of the profile.

\begin{figure}
\centering
\includegraphics[width=7cm, height=12cm]{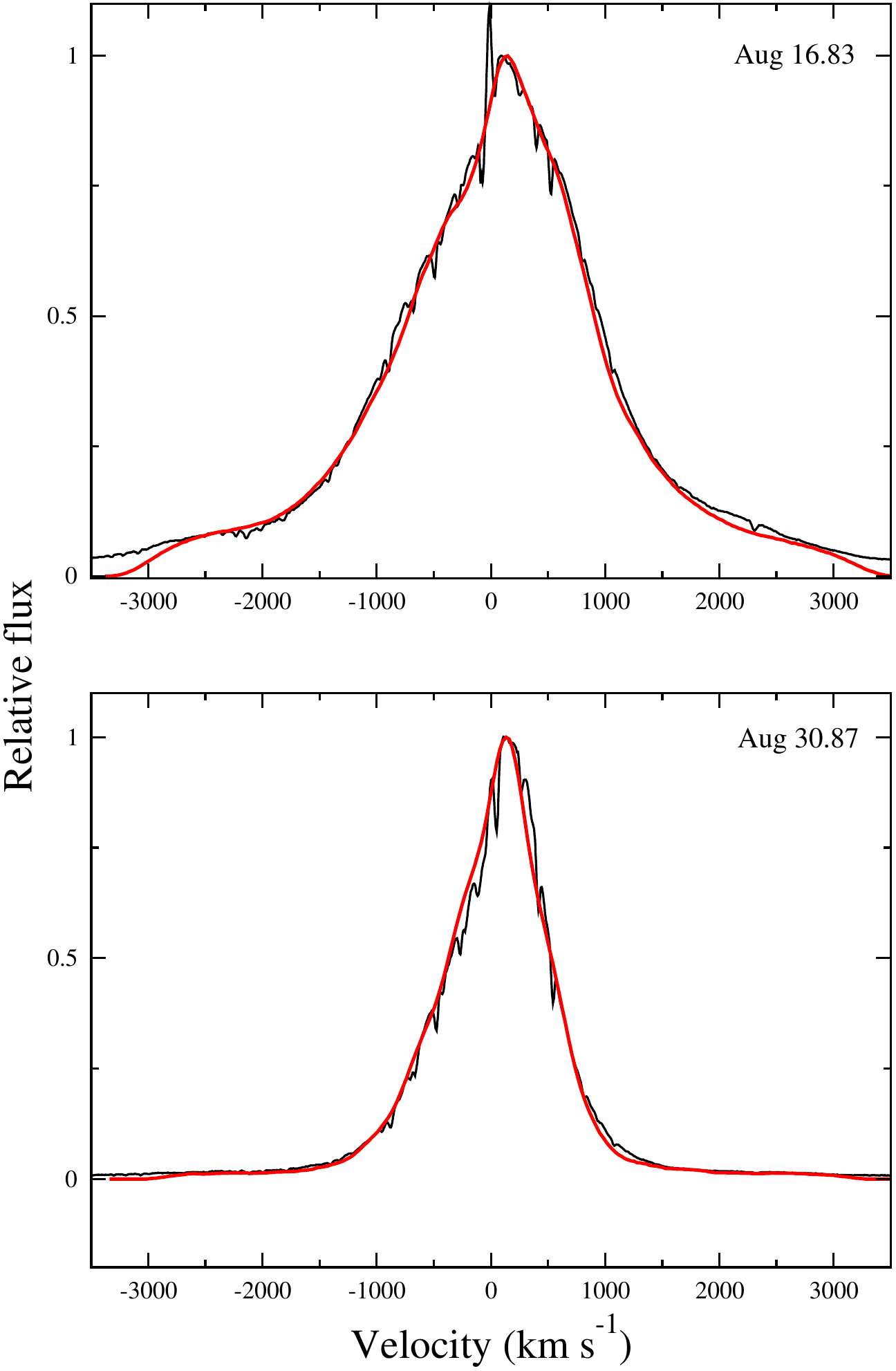}
\caption{The best-fitted \ion{H}{$\alpha$} profiles of RS Oph 2021 for the dates Aug. 16.83 (above row) and 30.87 (below) from 3D morpho-kinematic modeling. The observed profile is shown in black and the modelled profile is shown in red. \label{fig:shapeprofile}}
\end{figure}

\begin{figure}
\centering
\includegraphics[width=8cm, height=15cm]{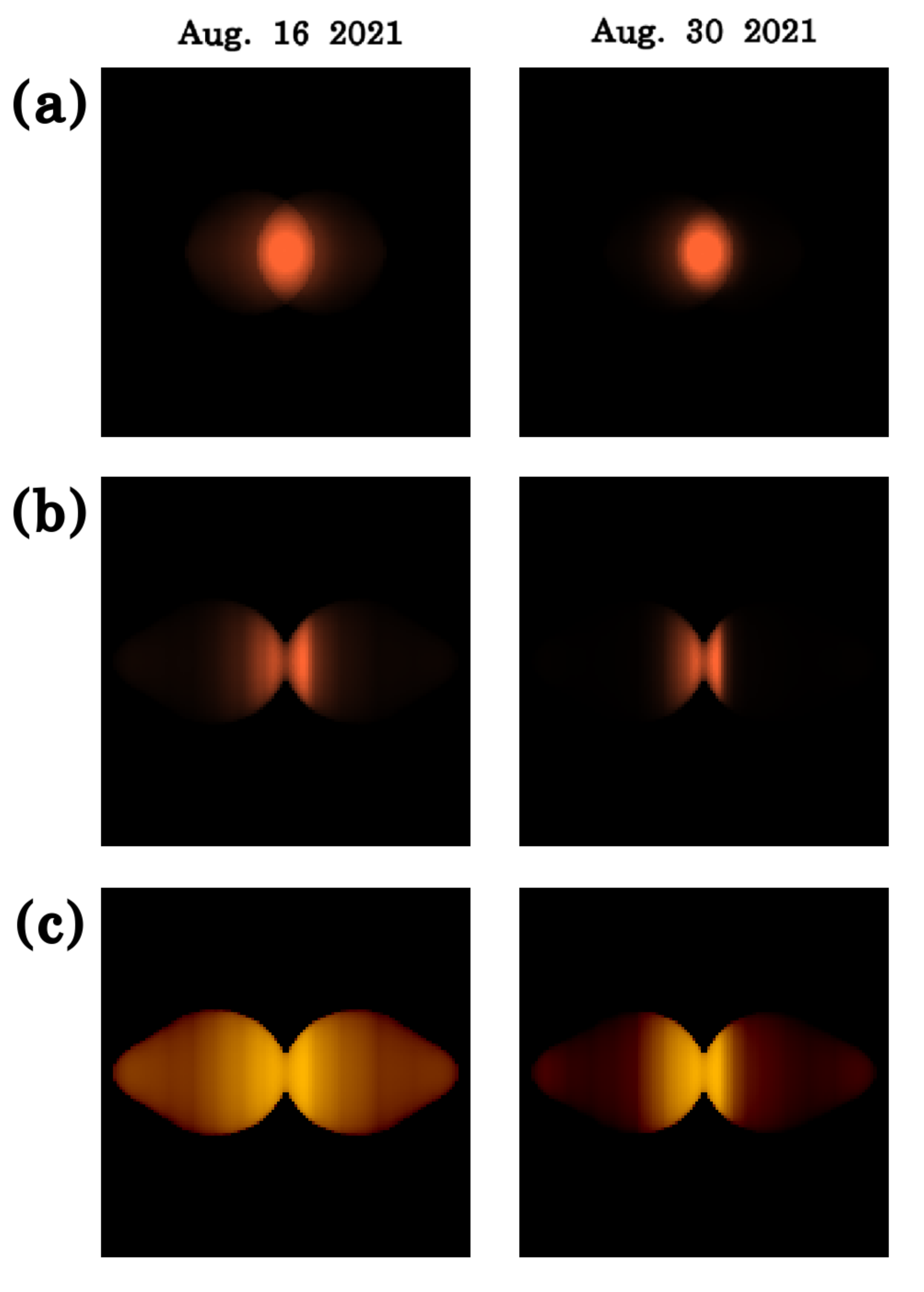}
\caption{The model images of the ejecta of RS Oph 2021 obtained from the best-fitted 3D morpho-kinematic models for Aug. 16.83 (left column) and 30.87 (right column), 2021. Panel (a) shows the image of the ejecta along the line-of-sight. The density distribution ($n=n(z)$) is shown in the side view of the ejecta in linear scale (Panel (b)) and in log scale (Panel (c)). \label{fig:shapeimage}}
\end{figure}

We obtained the final model that best fits both \ion{H}{$\alpha$} profiles on Aug. 16.83 and 30.87, 2021. The fits of the profiles are shown in Fig.~\ref{fig:shapeprofile}. The final model showed a bipolar morphology. The best-fitting inclination from our modelling was obtained as $i = 30^{\circ}$. The modelled images on both dates as viewed in the sky are shown in Fig.~\ref{fig:shapeimage}(a). We obtain a density profile varying along the bipolar axis $(n = n(z))$. Fig.~\ref{fig:shapeimage}(b) shows the variation of density distribution along the bipolar axis. Fig.~\ref{fig:shapeimage}(b) more clearly depicts the same in log scale. The maximum expansion velocity is obtained as 2000 and 1000 kms$^{-1}$, on the dates Aug 16.83 and 30.87, 2021, respectively. 

In this work, we mainly aimed for the basic physical structure of the nova ejecta. Hence, we did not attempt for a detailed optimization technique to minimize errors and obtain the final model. We used visual estimation to finalize our model. We narrowed down the model parameter range by noting the characteristic deviation of the modeled spectra from the observed spectra. By this inspection, an error of $\sim10{\%}$ can be attributed to the model parameters. 

\subsection{Model limitaions}
The low $\chi^2_{red}$ ($< 2$) values of the model in the present study indicates that the \textsc{cloudy} model generated spectra match well with the observed spectra. Even while our model reproduces a wide variety of observable effects, the phenomenology has certain limits, which we explain below. 

Our 3D morpho-kinematic modelling of the prominent \ion{H}{$\alpha$} emission lines indicates a non-spherical bipolar geometry of the ejecta (Section~\ref{sect:3D}). The previous studies of RS Oph during the 2006 outburst at X-ray, optical, IR, and radio wavelengths also suggested a bipolar structure of the ejected material. For simplicity, we assumed a homogeneous spherically symmetric geometry for \textsc{cloudy} modelling (Section~\ref{sect:cloudy}). We understand that a 1D code like \textsc{cloudy} is not well suited to representing such a complex environment. It executes the two-component model separately, although they are not separate in reality. We plan to incorporate a more detailed treatment of non-spherical geometry in a future work.

It may be noted that few parameters could be correlated, so it is extremely difficult to determine the uniqueness of our model. However, we have succeeded in reproducing a wide range of observational effects, and we believe that our model can explain several important aspects of the nova shell of RS Oph during its 2021 outburst.

Previous TNR hydrodynamics studies on WD for novae ejecta revealed that the ejecta density varies with ejecta radius according to a power law, $\rho \propto r^{-\alpha}$, with the value of $\alpha = 10$ for the early \enquote{fireball} phase \citep{1998MNRASStarrfield,1992ApJHauschildt,1995ApJHauschildt}. The temperature during the early fireball phase is around 15,000 K and it drops to $<$ 10,000 K with the adbiatic expansion of ejecta. The optical depth is reduced as a result of the expansion, making the deeper layers of the ejecta visible. The density profile of ejecta grows shallower at this stage, which is best modelled with a smaller value of $\alpha \sim 3$ \citep{1992ApJHauschildt}. For a constant mass-loss rate, this density profile corresponds to a velocity law of the form $v(r) \propto r^{\alpha -2}$ \citep{2001MNRAS.320..103S}. This small value of $\alpha$ leads to a very large geometrical extension of the novae atmospheres. The RS Oph system, on the other hand, is a symbiotic recurrent nova in which the primary WD orbits within the winds of a red giant companion. These systems create expanding shock waves, which sweep the winds of the red giant. Because of the constraints of 1D \textsc{cloudy} modelling, the influence of shock waves could not be examined in our basic phenomenological method. In a forthcoming work, we intend to add a more extensive investigation of nonspherical ejecta geometry and shock wave effects while modelling nova ejecta for RS Oph.

\section{Summary}\label{sect:summ}
We reported the evolution of the optical spectra of the 2021 outburst of the galactic recurrent nova RS Oph over about a month after the outburst. We found that the spectral evolution is similar to that of the previous outbursts. The nova reached maximum brightness on Aug 10.1 with the V band magnitude of 4.5. The decline timescales $t_2$ and $t_3$ are determined to be 2.77 days and 8.04 days, respectively. The distance of the system is determined to be $\sim 1.68$ kpc. The optical spectra in the early days showed sharp, narrow P-cygni emission features superimposed over the broad emission features, which later became narrower and sharper as the nova evolved. After the outburst, the ejected material was in the free expansion phase for almost 4 days, and then eventually the material moving outward with a very high velocity was obstructed by the surrounding winds of the red giant companion and generated a shock wave. The shock velocity propagates into the winds of the red giant and its velocity declines monotonically with time as $v\propto t^{-0.6}$. We used the 1D photoionization code \textsc{cloudy} to model the observed optical spectra of four epochs during the first month following the outburst. Our model shows the presence of a hot central ionising WD with a temperature of $4.16 \times 10^4$K, $4.89 \times 10^4$K, $6.45 \times 10^4$K, and $6.60 \times 10^4$K for epochs 1, 2, 3, and 4, respectively, with a roughly constant luminosity of $\sim 1.0 \times 10^{37}$erg s$^{-1}$. The estimated abundance from the best-fit \textsc{cloudy} models are: He/H $\sim 1.4 - 1.9$, N/H = $70 - 95$, O/H = $0.6 - 2.6$, and Fe/H $\sim 1.0 - 1.9$ for the ejecta during the first month after the outburst. We estimated the ejected hydrogen shell mass of the system in the range of $3.54 - 3.83 \times 10^{-6} M_{\odot}$. From the 3D morpho-kinematic modelling of the nova, we found a bipolar ejecta morphology. The inclination angle of the binary system was determined to be $i=30^{\circ}$.

\section*{acknowledgements}
The authors would like to express their gratitude to Prof. Aneurin Evans, the referee, for his insightful remarks and recommendations, which significantly improved the manuscript. The research work at the S. N. Bose National Centre for Basic Sciences is funded by the Department of Science and Technology, Government of India. We are grateful to all the observers at the ARAS database that made their observations available for public use. In particular, we are thankful to Olivier Thizy (OTH) for his spectroscopic observations. We also thank the AAVSO database, for using their photometric data. 
\section*{Data availability}
Spectra used in the paper are available in the ARAS Database. \url{https://aras-database.github.io/database/rsoph.html}
The photometric data used to generate light curve (Fig.~\ref{fig:lightcurve}) are available in the AAVSO Database. \url{https://www.aavso.org/LCGv2/}
\bibliographystyle{mnras}
\bibliography{references}

\begin{thebibliography}{}
\makeatletter
\relax
\def\mn@urlcharsother{\let\do\@makeother \do\$\do\&\do\#\do\^\do\_\do\%\do\~}
\def\mn@doi{\begingroup\mn@urlcharsother \@ifnextchar [ {\mn@doi@}
  {\mn@doi@[]}}
\def\mn@doi@[#1]#2{\def\@tempa{#1}\ifx\@tempa\@empty \href
  {http://dx.doi.org/#2} {doi:#2}\else \href {http://dx.doi.org/#2} {#1}\fi
  \endgroup}
\def\mn@eprint#1#2{\mn@eprint@#1:#2::\@nil}
\def\mn@eprint@arXiv#1{\href {http://arxiv.org/abs/#1} {{\tt arXiv:#1}}}
\def\mn@eprint@dblp#1{\href {http://dblp.uni-trier.de/rec/bibtex/#1.xml}
  {dblp:#1}}
\def\mn@eprint@#1:#2:#3:#4\@nil{\def\@tempa {#1}\def\@tempb {#2}\def\@tempc
  {#3}\ifx \@tempc \@empty \let \@tempc \@tempb \let \@tempb \@tempa \fi \ifx
  \@tempb \@empty \def\@tempb {arXiv}\fi \@ifundefined
  {mn@eprint@\@tempb}{\@tempb:\@tempc}{\expandafter \expandafter \csname
  mn@eprint@\@tempb\endcsname \expandafter{\@tempc}}}

\bibitem[\protect\citeauthoryear{{Anupama} \& {Prabhu}}{{Anupama} \&
  {Prabhu}}{1989}]{1989Anupama}
{Anupama} G.~C.,  {Prabhu} T.~P.,  1989, \mn@doi [Journal of Astrophysics and
  Astronomy] {10.1007/BF02714998}, \href
  {https://ui.adsabs.harvard.edu/abs/1989JApA...10..237A} {10, 237}

\bibitem[\protect\citeauthoryear{{Barbon}, {Mammano}  \& {Rosino}}{{Barbon}
  et~al.}{1969}]{1969CoKon..65..257B}
{Barbon} R.,  {Mammano} A.,   {Rosino} L.,  1969, Commmunications of the
  Konkoly Observatory Hungary, \href
  {https://ui.adsabs.harvard.edu/abs/1969CoKon..65..257B} {65, 257}

\bibitem[\protect\citeauthoryear{{Barry}, {Mukai}, {Sokoloski}, {Danchi},
  {Hachisu}, {Evans}, {Gehrz}  \& {Mikolajewska}}{{Barry}
  et~al.}{2008}]{2008ASPC..401...52B}
{Barry} R.~K.,  {Mukai} K.,  {Sokoloski} J.~L.,  {Danchi} W.~C.,  {Hachisu} I.,
   {Evans} A.,  {Gehrz} R.,   {Mikolajewska} J.,  2008, in {Evans} A.,  {Bode}
  M.~F.,  {O'Brien} T.~J.,   {Darnley} M.~J.,  eds,  Astronomical Society of
  the Pacific Conference Series Vol. 401, RS Ophiuchi (2006) and the Recurrent
  Nova Phenomenon. p.~52

\bibitem[\protect\citeauthoryear{{Bath} \& {Shaviv}}{{Bath} \&
  {Shaviv}}{1976a}]{1976MNRAS.175..305B}
{Bath} G.~T.,  {Shaviv} G.,  1976a, \mn@doi [\mnras] {10.1093/mnras/175.2.305},
  \href {https://ui.adsabs.harvard.edu/abs/1976MNRAS.175..305B} {175, 305}

\bibitem[\protect\citeauthoryear{{Bath} \& {Shaviv}}{{Bath} \&
  {Shaviv}}{1976b}]{1976MNRASBath&Shaviv}
{Bath} G.~T.,  {Shaviv} G.,  1976b, \mn@doi [\mnras] {10.1093/mnras/175.2.305},
  \href {https://ui.adsabs.harvard.edu/abs/1976MNRAS.175..305B} {175, 305}

\bibitem[\protect\citeauthoryear{{Bode}}{{Bode}}{1987}]{1987Bode}
{Bode} M.~F.,  1987, {RS Ophiuchi (1985) and the recurrent nova phenomenon.
  Proceedings of the Manchester Conference, held at Manchester, UK, 16 - 18
  December 1985.}

\bibitem[\protect\citeauthoryear{{Bode} \& {Evans}}{{Bode} \&
  {Evans}}{2008}]{2008clno.bookBode}
{Bode} M.~F.,  {Evans} A.,  2008, {Classical Novae}.
 Vol. 43

\bibitem[\protect\citeauthoryear{{Bode} \& {Kahn}}{{Bode} \&
  {Kahn}}{1985}]{1985BodeKahn}
{Bode} M.~F.,  {Kahn} F.~D.,  1985, \mn@doi [\mnras] {10.1093/mnras/217.1.205},
  \href {https://ui.adsabs.harvard.edu/abs/1985MNRAS.217..205B} {217, 205}

\bibitem[\protect\citeauthoryear{{Bode} et~al.,}{{Bode}
  et~al.}{2006}]{2006BodeRSOphUV}
{Bode} M.~F.,  et~al., 2006, \mn@doi [\apj] {10.1086/507980}, \href
  {https://ui.adsabs.harvard.edu/abs/2006ApJ...652..629B} {652, 629}

\bibitem[\protect\citeauthoryear{{Booth}, {Mohamed}  \&
  {Podsiadlowski}}{{Booth} et~al.}{2016}]{2016Booth}
{Booth} R.~A.,  {Mohamed} S.,   {Podsiadlowski} P.,  2016, \mn@doi [\mnras]
  {10.1093/mnras/stw001}, \href
  {https://ui.adsabs.harvard.edu/abs/2016MNRAS.457..822B} {457, 822}

\bibitem[\protect\citeauthoryear{{Brandi}, {Quiroga}, {Miko{\l}ajewska},
  {Ferrer}  \& {Garc{\'\i}a}}{{Brandi} et~al.}{2009}]{2009brandi}
{Brandi} E.,  {Quiroga} C.,  {Miko{\l}ajewska} J.,  {Ferrer} O.~E.,
  {Garc{\'\i}a} L.~G.,  2009, \mn@doi [\aap] {10.1051/0004-6361/200811417},
  \href {https://ui.adsabs.harvard.edu/abs/2009A&A...497..815B} {497, 815}

\bibitem[\protect\citeauthoryear{{Cassatella}, {Hassall}, {Harris}  \&
  {Snijders}}{{Cassatella} et~al.}{1985}]{1985cassatella}
{Cassatella} A.,  {Hassall} B.~J.~M.,  {Harris} A.,   {Snijders} M.~A.~J.,
  1985, in {Burke} W.~R.,  ed., Recent Results on Cataclysmic Variables. The
  Importance of IUE and Exosat Results on Cataclysmic Variables and Low-Mass
  X-Ray Binaries. p.~281

\bibitem[\protect\citeauthoryear{{Cheung}, {Ciprini}  \& {Johnson}}{{Cheung}
  et~al.}{2021}]{2021ATelFermi-LAT}
{Cheung} C.~C.,  {Ciprini} S.,   {Johnson} T.~J.,  2021, The Astronomer's
  Telegram, \href {https://ui.adsabs.harvard.edu/abs/2021ATel14834....1C}
  {14834, 1}

\bibitem[\protect\citeauthoryear{{Contini}, {Orio}  \& {Prialnik}}{{Contini}
  et~al.}{1995}]{1995MNRAScontini}
{Contini} M.,  {Orio} M.,   {Prialnik} D.,  1995, \mn@doi [\mnras]
  {10.1093/mnras/275.1.195}, \href
  {https://ui.adsabs.harvard.edu/abs/1995MNRAS.275..195C} {275, 195}

\bibitem[\protect\citeauthoryear{{Das} \& {Mondal}}{{Das} \&
  {Mondal}}{2015}]{2015DasMondal}
{Das} R.,  {Mondal} A.,  2015, \mn@doi [\na] {10.1016/j.newast.2015.02.004},
  \href {https://ui.adsabs.harvard.edu/abs/2015NewA...39...19D} {39, 19}

\bibitem[\protect\citeauthoryear{{Das}, {Banerjee}  \& {Ashok}}{{Das}
  et~al.}{2006}]{2006DasRSOph}
{Das} R.,  {Banerjee} D. P.~K.,   {Ashok} N.~M.,  2006, \mn@doi [\apjl]
  {10.1086/510674}, \href
  {https://ui.adsabs.harvard.edu/abs/2006ApJ...653L.141D} {653, L141}

\bibitem[\protect\citeauthoryear{{Dobrzycka} \& {Kenyon}}{{Dobrzycka} \&
  {Kenyon}}{1994}]{1994AJDobrzycka}
{Dobrzycka} D.,  {Kenyon} S.~J.,  1994, \mn@doi [\aj] {10.1086/117238}, \href
  {https://ui.adsabs.harvard.edu/abs/1994AJ....108.2259D} {108, 2259}

\bibitem[\protect\citeauthoryear{{Drake} et~al.,}{{Drake}
  et~al.}{2009}]{2009ApJDrake}
{Drake} J.~J.,  et~al., 2009, \mn@doi [\apj] {10.1088/0004-637X/691/1/418},
  \href {https://ui.adsabs.harvard.edu/abs/2009ApJ...691..418D} {691, 418}

\bibitem[\protect\citeauthoryear{{Enoto} et~al.,}{{Enoto}
  et~al.}{2021}]{2021ATelNicer}
{Enoto} T.,  et~al., 2021, The Astronomer's Telegram, \href
  {https://ui.adsabs.harvard.edu/abs/2021ATel14850....1E} {14850, 1}

\bibitem[\protect\citeauthoryear{{Evans}, {Callus}, {Albinson}, {Whitelock},
  {Glass}, {Carter}  \& {Roberts}}{{Evans} et~al.}{1988}]{1988EvansIR}
{Evans} A.,  {Callus} C.~M.,  {Albinson} J.~S.,  {Whitelock} P.~A.,  {Glass}
  I.~S.,  {Carter} B.,   {Roberts} G.,  1988, \mn@doi [\mnras]
  {10.1093/mnras/234.3.755}, \href
  {https://ui.adsabs.harvard.edu/abs/1988MNRAS.234..755E} {234, 755}

\bibitem[\protect\citeauthoryear{{Evans} et~al.,}{{Evans}
  et~al.}{2007a}]{2007MNRASEvans}
{Evans} A.,  et~al., 2007a, \mn@doi [\mnras]
  {10.1111/j.1745-3933.2006.00252.x}, \href
  {https://ui.adsabs.harvard.edu/abs/2007MNRAS.374L...1E} {374, L1}

\bibitem[\protect\citeauthoryear{{Evans} et~al.,}{{Evans}
  et~al.}{2007b}]{2007ApJEvans}
{Evans} A.,  et~al., 2007b, \mn@doi [\apjl] {10.1086/519924}, \href
  {https://ui.adsabs.harvard.edu/abs/2007ApJ...663L..29E} {663, L29}

\bibitem[\protect\citeauthoryear{{Ferland} et~al.,}{{Ferland}
  et~al.}{2013}]{2013RMxAA..49..137F}
{Ferland} G.~J.,  et~al., 2013, \rmxaa, \href
  {https://ui.adsabs.harvard.edu/abs/2013RMxAA..49..137F} {49, 137}

\bibitem[\protect\citeauthoryear{{Ferland} et~al.,}{{Ferland}
  et~al.}{2017}]{2017RMxAA..53..385F}
{Ferland} G.~J.,  et~al., 2017, \rmxaa, \href
  {https://ui.adsabs.harvard.edu/abs/2017RMxAA..53..385F} {53, 385}

\bibitem[\protect\citeauthoryear{{Ferrigno} et~al.,}{{Ferrigno}
  et~al.}{2021}]{2021ATelINTEGRAL}
{Ferrigno} C.,  et~al., 2021, The Astronomer's Telegram, \href
  {https://ui.adsabs.harvard.edu/abs/2021ATel14855....1F} {14855, 1}

\bibitem[\protect\citeauthoryear{{Gehrz}, {Truran}, {Williams}  \&
  {Starrfield}}{{Gehrz} et~al.}{1998}]{1998Gehrz}
{Gehrz} R.~D.,  {Truran} J.~W.,  {Williams} R.~E.,   {Starrfield} S.,  1998,
  \mn@doi [\pasp] {10.1086/316107}, \href
  {https://ui.adsabs.harvard.edu/abs/1998PASP..110....3G} {110, 3}

\bibitem[\protect\citeauthoryear{{Grevesse}, {Asplund}, {Sauval}  \&
  {Scott}}{{Grevesse} et~al.}{2010}]{2010Ap&SS.328..179G}
{Grevesse} N.,  {Asplund} M.,  {Sauval} A.~J.,   {Scott} P.,  2010, \mn@doi
  [\apss] {10.1007/s10509-010-0288-z}, \href
  {https://ui.adsabs.harvard.edu/abs/2010Ap&SS.328..179G} {328, 179}

\bibitem[\protect\citeauthoryear{{Hachisu} \& {Kato}}{{Hachisu} \&
  {Kato}}{2001}]{2001ApJHachisu}
{Hachisu} I.,  {Kato} M.,  2001, \mn@doi [\apj] {10.1086/321601}, \href
  {https://ui.adsabs.harvard.edu/abs/2001ApJ...558..323H} {558, 323}

\bibitem[\protect\citeauthoryear{{Hachisu}, {Kato}  \& {Luna}}{{Hachisu}
  et~al.}{2007}]{2007ApJHachisu}
{Hachisu} I.,  {Kato} M.,   {Luna} G. J.~M.,  2007, \mn@doi [\apjl]
  {10.1086/516838}, \href
  {https://ui.adsabs.harvard.edu/abs/2007ApJ...659L.153H} {659, L153}

\bibitem[\protect\citeauthoryear{{Hauschildt}, {Wehrse}, {Starrfield}  \&
  {Shaviv}}{{Hauschildt} et~al.}{1992}]{1992ApJHauschildt}
{Hauschildt} P.~H.,  {Wehrse} R.,  {Starrfield} S.,   {Shaviv} G.,  1992,
  \mn@doi [\apj] {10.1086/171507}, \href
  {https://ui.adsabs.harvard.edu/abs/1992ApJ...393..307H} {393, 307}

\bibitem[\protect\citeauthoryear{{Hauschildt}, {Starrfield}, {Shore}, {Allard}
  \& {Baron}}{{Hauschildt} et~al.}{1995}]{1995ApJHauschildt}
{Hauschildt} P.~H.,  {Starrfield} S.,  {Shore} S.~N.,  {Allard} F.,   {Baron}
  E.,  1995, \mn@doi [\apj] {10.1086/175921}, \href
  {https://ui.adsabs.harvard.edu/abs/1995ApJ...447..829H} {447, 829}

\bibitem[\protect\citeauthoryear{{Hauschildt}, {Shore}, {Schwarz}, {Baron},
  {Starrfield}  \& {Allard}}{{Hauschildt} et~al.}{1997}]{1997ApJHauschildt}
{Hauschildt} P.~H.,  {Shore} S.~N.,  {Schwarz} G.~J.,  {Baron} E.,
  {Starrfield} S.,   {Allard} F.,  1997, \mn@doi [\apj] {10.1086/304904}, \href
  {https://ui.adsabs.harvard.edu/abs/1997ApJ...490..803H} {490, 803}

\bibitem[\protect\citeauthoryear{{Helton} et~al.,}{{Helton}
  et~al.}{2010}]{2010AJ....140.1347H}
{Helton} L.~A.,  et~al., 2010, \mn@doi [\aj] {10.1088/0004-6256/140/5/1347},
  \href {https://ui.adsabs.harvard.edu/abs/2010AJ....140.1347H} {140, 1347}

\bibitem[\protect\citeauthoryear{{Hjellming}, {van Gorkom}, {Taylor},
  {Sequist}, {Padin}, {Davis}  \& {Bode}}{{Hjellming}
  et~al.}{1986}]{1986ApJ...305L..71H}
{Hjellming} R.~M.,  {van Gorkom} J.~H.,  {Taylor} A.~R.,  {Sequist} E.~R.,
  {Padin} S.,  {Davis} R.~J.,   {Bode} M.~F.,  1986, \mn@doi [\apjl]
  {10.1086/184687}, \href
  {https://ui.adsabs.harvard.edu/abs/1986ApJ...305L..71H} {305, L71}

\bibitem[\protect\citeauthoryear{{Iijima}}{{Iijima}}{2009}]{Iijima09}
{Iijima} T.,  2009, \mn@doi [\aap] {10.1051/0004-6361/200811436}, \href
  {https://ui.adsabs.harvard.edu/abs/2009A&A...505..287I} {505, 287}

\bibitem[\protect\citeauthoryear{{Ikeda} \& {Tamura}}{{Ikeda} \&
  {Tamura}}{2004}]{2004PASJIkeda}
{Ikeda} Y.,  {Tamura} S.,  2004, \mn@doi [\pasj] {10.1093/pasj/56.2.353}, \href
  {https://ui.adsabs.harvard.edu/abs/2004PASJ...56..353I} {56, 353}

\bibitem[\protect\citeauthoryear{{Jos{\'e}}}{{Jos{\'e}}}{2012}]{2012BASIJose}
{Jos{\'e}} J.,  2012, Bulletin of the Astronomical Society of India, \href
  {https://ui.adsabs.harvard.edu/abs/2012BASI...40..443J} {40, 443}

\bibitem[\protect\citeauthoryear{{Jos{\'e}}, {Shore}  \& {Casanova}}{{Jos{\'e}}
  et~al.}{2020}]{2020A&AJose}
{Jos{\'e}} J.,  {Shore} S.~N.,   {Casanova} J.,  2020, \mn@doi [\aap]
  {10.1051/0004-6361/201936893}, \href
  {https://ui.adsabs.harvard.edu/abs/2020A&A...634A...5J} {634, A5}

\bibitem[\protect\citeauthoryear{{Krautter}}{{Krautter}}{2008}]{2008Krautter}
{Krautter} J.,  2008, in {Evans} A.,  {Bode} M.~F.,  {O'Brien} T.~J.,
  {Darnley} M.~J.,  eds,  Astronomical Society of the Pacific Conference Series
  Vol. 401, RS Ophiuchi (2006) and the Recurrent Nova Phenomenon. p.~139

\bibitem[\protect\citeauthoryear{{Livio} \& {Truran}}{{Livio} \&
  {Truran}}{1994}]{1994ApJLivio}
{Livio} M.,  {Truran} J.~W.,  1994, \mn@doi [\apj] {10.1086/174024}, \href
  {https://ui.adsabs.harvard.edu/abs/1994ApJ...425..797L} {425, 797}

\bibitem[\protect\citeauthoryear{{Miko{\l}ajewska} \&
  {Shara}}{{Miko{\l}ajewska} \& {Shara}}{2017}]{2017ApJ...847...99M}
{Miko{\l}ajewska} J.,  {Shara} M.~M.,  2017, \mn@doi [\apj]
  {10.3847/1538-4357/aa87b6}, \href
  {https://ui.adsabs.harvard.edu/abs/2017ApJ...847...99M} {847, 99}

\bibitem[\protect\citeauthoryear{{Mikolajewska}, {Aydi}, {Buckley}, {Galan}  \&
  {Orio}}{{Mikolajewska} et~al.}{2021a}]{2021ATelSALT}
{Mikolajewska} J.,  {Aydi} E.,  {Buckley} D.,  {Galan} C.,   {Orio} M.,  2021a,
  The Astronomer's Telegram, \href
  {https://ui.adsabs.harvard.edu/abs/2021ATel14852....1M} {14852, 1}

\bibitem[\protect\citeauthoryear{{Mikolajewska}, {Aydi}, {Buckley}, {Galan}  \&
  {Orio}}{{Mikolajewska} et~al.}{2021b}]{Mikolajewska2021}
{Mikolajewska} J.,  {Aydi} E.,  {Buckley} D.,  {Galan} C.,   {Orio} M.,  2021b,
  The Astronomer's Telegram, \href
  {https://ui.adsabs.harvard.edu/abs/2021ATel14852....1M} {14852, 1}

\bibitem[\protect\citeauthoryear{{Mondal}, {Anupama}, {Kamath}, {Das},
  {Selvakumar}  \& {Mondal}}{{Mondal} et~al.}{2018}]{2018Mondal}
{Mondal} A.,  {Anupama} G.~C.,  {Kamath} U.~S.,  {Das} R.,  {Selvakumar} G.,
  {Mondal} S.,  2018, \mn@doi [\mnras] {10.1093/mnras/stx2988}, \href
  {https://ui.adsabs.harvard.edu/abs/2018MNRAS.474.4211M} {474, 4211}

\bibitem[\protect\citeauthoryear{{Mondal}, {Das}, {Anupama}  \&
  {Mondal}}{{Mondal} et~al.}{2020}]{2020MNRASMondal}
{Mondal} A.,  {Das} R.,  {Anupama} G.~C.,   {Mondal} S.,  2020, \mn@doi
  [\mnras] {10.1093/mnras/stz3570}, \href
  {https://ui.adsabs.harvard.edu/abs/2020MNRAS.492.2326M} {492, 2326}

\bibitem[\protect\citeauthoryear{{Montez}, {Luna}, {Mukai}, {Sokoloski}  \&
  {Kastner}}{{Montez} et~al.}{2022}]{2022ApJMontez}
{Montez} R.,  {Luna} G.~J.~M.,  {Mukai} K.,  {Sokoloski} J.~L.,   {Kastner}
  J.~H.,  2022, \mn@doi [\apj] {10.3847/1538-4357/ac4583}, \href
  {https://ui.adsabs.harvard.edu/abs/2022ApJ...926..100M} {926, 100}

\bibitem[\protect\citeauthoryear{{Munari} \& {Valisa}}{{Munari} \&
  {Valisa}}{2021a}]{2021Munari}
{Munari} U.,  {Valisa} P.,  2021a, arXiv e-prints, \href
  {https://ui.adsabs.harvard.edu/abs/2021arXiv210901101M} {p. arXiv:2109.01101}

\bibitem[\protect\citeauthoryear{{Munari} \& {Valisa}}{{Munari} \&
  {Valisa}}{2021b}]{Munari2021}
{Munari} U.,  {Valisa} P.,  2021b, The Astronomer's Telegram, \href
  {https://ui.adsabs.harvard.edu/abs/2021ATel14840....1M} {14840, 1}

\bibitem[\protect\citeauthoryear{{Munari} \& {Valisa}}{{Munari} \&
  {Valisa}}{2022}]{2022arXivMunari}
{Munari} U.,  {Valisa} P.,  2022, arXiv e-prints, \href
  {https://ui.adsabs.harvard.edu/abs/2022arXiv220301378M} {p. arXiv:2203.01378}

\bibitem[\protect\citeauthoryear{{Munari} et~al.,}{{Munari}
  et~al.}{2007}]{2007BaltA..16...46M}
{Munari} U.,  et~al., 2007, Baltic Astronomy, \href
  {https://ui.adsabs.harvard.edu/abs/2007BaltA..16...46M} {16, 46}

\bibitem[\protect\citeauthoryear{{Ness} et~al.,}{{Ness}
  et~al.}{2009}]{2009AJNess}
{Ness} J.~U.,  et~al., 2009, \mn@doi [\aj] {10.1088/0004-6256/137/2/3414},
  \href {https://ui.adsabs.harvard.edu/abs/2009AJ....137.3414N} {137, 3414}

\bibitem[\protect\citeauthoryear{{Nikolov} \& {Luna}}{{Nikolov} \&
  {Luna}}{2021}]{2021ATelpolarisation}
{Nikolov} Y.,  {Luna} G.~J.~M.,  2021, The Astronomer's Telegram, \href
  {https://ui.adsabs.harvard.edu/abs/2021ATel14863....1N} {14863, 1}

\bibitem[\protect\citeauthoryear{{O'Brien} et~al.,}{{O'Brien}
  et~al.}{2006}]{2006NaturO'brien}
{O'Brien} T.~J.,  et~al., 2006, \mn@doi [\nat] {10.1038/nature04949}, \href
  {https://ui.adsabs.harvard.edu/abs/2006Natur.442..279O} {442, 279}

\bibitem[\protect\citeauthoryear{{Oppenheimer} \& {Mattei}}{{Oppenheimer} \&
  {Mattei}}{1993}]{1993AAS...183.5503O}
{Oppenheimer} B.,  {Mattei} J.~A.,  1993, in American Astronomical Society
  Meeting Abstracts. p. 55.03

\bibitem[\protect\citeauthoryear{{Orio}, {Behar}, {Drake}, {Mikolajewska},
  {Ness}  \& {Ospina}}{{Orio} et~al.}{2021a}]{2021ATelorio2}
{Orio} M.,  {Behar} E.,  {Drake} J.,  {Mikolajewska} J.,  {Ness} J.~U.,
  {Ospina} N.,  2021a, The Astronomer's Telegram, \href
  {https://ui.adsabs.harvard.edu/abs/2021ATel14906....1O} {14906, 1}

\bibitem[\protect\citeauthoryear{{Orio} et~al.,}{{Orio}
  et~al.}{2021b}]{2021ATelorio}
{Orio} M.,  et~al., 2021b, The Astronomer's Telegram, \href
  {https://ui.adsabs.harvard.edu/abs/2021ATel14926....1O} {14926, 1}

\bibitem[\protect\citeauthoryear{{Orlando}, {Drake}  \& {Laming}}{{Orlando}
  et~al.}{2009}]{2009A&Aorlando}
{Orlando} S.,  {Drake} J.~J.,   {Laming} J.~M.,  2009, \mn@doi [\aap]
  {10.1051/0004-6361:200810109}, \href
  {https://ui.adsabs.harvard.edu/abs/2009A&A...493.1049O} {493, 1049}

\bibitem[\protect\citeauthoryear{{Page}}{{Page}}{2021}]{2021ATelPage}
{Page} K.~L.,  2021, The Astronomer's Telegram, \href
  {https://ui.adsabs.harvard.edu/abs/2021ATel14894....1P} {14894, 1}

\bibitem[\protect\citeauthoryear{{Page}, {Osborne}  \& {Aydi}}{{Page}
  et~al.}{2021}]{2021ATelSwift-XRT}
{Page} K.~L.,  {Osborne} J.~P.,   {Aydi} E.,  2021, The Astronomer's Telegram,
  \href {https://ui.adsabs.harvard.edu/abs/2021ATel14848....1P} {14848, 1}

\bibitem[\protect\citeauthoryear{Pandey, Das, Shaw  \& Mondal}{Pandey
  et~al.}{2022}]{Pandey_2022}
Pandey R.,  Das R.,  Shaw G.,   Mondal S.,  2022, \mn@doi [The Astrophysical
  Journal] {10.3847/1538-4357/ac36dc}, 925, 187

\bibitem[\protect\citeauthoryear{{Pavana}, {Anche}, {Anupama}, {Ramaprakash}
  \& {Selvakumar}}{{Pavana} et~al.}{2019}]{2019A&A...622A.126P}
{Pavana} M.,  {Anche} R.~M.,  {Anupama} G.~C.,  {Ramaprakash} A.~N.,
  {Selvakumar} G.,  2019, \mn@doi [\aap] {10.1051/0004-6361/201833728}, \href
  {https://ui.adsabs.harvard.edu/abs/2019A&A...622A.126P} {622, A126}

\bibitem[\protect\citeauthoryear{{Pavlenko} et~al.,}{{Pavlenko}
  et~al.}{2008}]{2008A&APavlenko}
{Pavlenko} Y.~V.,  et~al., 2008, \mn@doi [\aap] {10.1051/0004-6361:20078622},
  \href {https://ui.adsabs.harvard.edu/abs/2008A&A...485..541P} {485, 541}

\bibitem[\protect\citeauthoryear{{Ribeiro} et~al.,}{{Ribeiro}
  et~al.}{2009}]{2009ApJRibeiro}
{Ribeiro} V.~A.~R.~M.,  et~al., 2009, \mn@doi [\apj]
  {10.1088/0004-637X/703/2/1955}, \href
  {https://ui.adsabs.harvard.edu/abs/2009ApJ...703.1955R} {703, 1955}

\bibitem[\protect\citeauthoryear{{Ribeiro}, {Darnley}, {Bode}, {Munari},
  {Harman}, {Steele}  \& {Meaburn}}{{Ribeiro} et~al.}{2011}]{2011MNRASRibeiro}
{Ribeiro} V.~A.~R.~M.,  {Darnley} M.~J.,  {Bode} M.~F.,  {Munari} U.,  {Harman}
  D.~J.,  {Steele} I.~A.,   {Meaburn} J.,  2011, \mn@doi [\mnras]
  {10.1111/j.1365-2966.2010.18006.x}, \href
  {https://ui.adsabs.harvard.edu/abs/2011MNRAS.412.1701R} {412, 1701}

\bibitem[\protect\citeauthoryear{{Rosino}}{{Rosino}}{1987}]{1987rorn.conf....1R}
{Rosino} L.,  1987, in {Bode} M.~F.,  ed., RS Ophiuchi (1985) and the Recurrent
  Nova Phenomenon. p.~1

\bibitem[\protect\citeauthoryear{{Rosino} \& {Iijima}}{{Rosino} \&
  {Iijima}}{1987}]{1987rosino}
{Rosino} L.,  {Iijima} T.,  1987, in {Bode} M.~F.,  ed., RS Ophiuchi (1985) and
  the Recurrent Nova Phenomenon. p.~27

\bibitem[\protect\citeauthoryear{{Rout}, {Srivastava}, {Banerjee}, {Vadawale},
  {Joshi}  \& {Kumar}}{{Rout} et~al.}{2021}]{2021ATelAstrosat}
{Rout} S.~K.,  {Srivastava} M.~K.,  {Banerjee} D. P.~K.,  {Vadawale} S.,
  {Joshi} V.,   {Kumar} V.,  2021, The Astronomer's Telegram, \href
  {https://ui.adsabs.harvard.edu/abs/2021ATel14882....1R} {14882, 1}

\bibitem[\protect\citeauthoryear{{Rudy}, {Erwin}, {Rossano}  \&
  {Puetter}}{{Rudy} et~al.}{1991}]{1991ApJRudy}
{Rudy} R.,  {Erwin} P.,  {Rossano} G.~S.,   {Puetter} R.~C.,  1991, \mn@doi
  [\apj] {10.1086/170792}, \href
  {https://ui.adsabs.harvard.edu/abs/1991ApJ...383..344R} {383, 344}

\bibitem[\protect\citeauthoryear{{Schaefer}}{{Schaefer}}{2004}]{2004Schaefer}
{Schaefer} B.~E.,  2004, \iaucirc, \href
  {https://ui.adsabs.harvard.edu/abs/2004IAUC.8396....2S} {8396, 2}

\bibitem[\protect\citeauthoryear{{Schaefer}}{{Schaefer}}{2010}]{2010Schaeffer}
{Schaefer} B.~E.,  2010, \mn@doi [\apjs] {10.1088/0067-0049/187/2/275}, \href
  {https://ui.adsabs.harvard.edu/abs/2010ApJS..187..275S} {187, 275}

\bibitem[\protect\citeauthoryear{{Schwarz}}{{Schwarz}}{2002}]{2002ApJ...577..940S}
{Schwarz} G.~J.,  2002, \mn@doi [\apj] {10.1086/342234}, \href
  {https://ui.adsabs.harvard.edu/abs/2002ApJ...577..940S} {577, 940}

\bibitem[\protect\citeauthoryear{{Schwarz}, {Starrfield}, {Shore}  \&
  {Hauschildt}}{{Schwarz} et~al.}{1997}]{1997MNRAS.290...75S}
{Schwarz} G.~J.,  {Starrfield} S.,  {Shore} S.~N.,   {Hauschildt} P.~H.,  1997,
  \mn@doi [\mnras] {10.1093/mnras/290.1.75}, \href
  {https://ui.adsabs.harvard.edu/abs/1997MNRAS.290...75S} {290, 75}

\bibitem[\protect\citeauthoryear{{Schwarz}, {Shore}, {Starrfield},
  {Hauschildt}, {Della Valle}  \& {Baron}}{{Schwarz}
  et~al.}{2001}]{2001MNRAS.320..103S}
{Schwarz} G.~J.,  {Shore} S.~N.,  {Starrfield} S.,  {Hauschildt} P.~H.,  {Della
  Valle} M.,   {Baron} E.,  2001, \mn@doi [\mnras]
  {10.1046/j.1365-8711.2001.03960.x}, \href
  {https://ui.adsabs.harvard.edu/abs/2001MNRAS.320..103S} {320, 103}

\bibitem[\protect\citeauthoryear{{Schwarz}, {Shore}, {Starrfield}  \&
  {Vanlandingham}}{{Schwarz} et~al.}{2007}]{2007ApJ...657..453S}
{Schwarz} G.~J.,  {Shore} S.~N.,  {Starrfield} S.,   {Vanlandingham} K.~M.,
  2007, \mn@doi [\apj] {10.1086/510661}, \href
  {https://ui.adsabs.harvard.edu/abs/2007ApJ...657..453S} {657, 453}

\bibitem[\protect\citeauthoryear{{Shidatsu} et~al.,}{{Shidatsu}
  et~al.}{2021}]{2021ATelMAXI}
{Shidatsu} M.,  et~al., 2021, The Astronomer's Telegram, \href
  {https://ui.adsabs.harvard.edu/abs/2021ATel14846....1S} {14846, 1}

\bibitem[\protect\citeauthoryear{{Shore}}{{Shore}}{2012}]{2012BASIShore}
{Shore} S.~N.,  2012, Bulletin of the Astronomical Society of India, \href
  {https://ui.adsabs.harvard.edu/abs/2012BASI...40..185S} {40, 185}

\bibitem[\protect\citeauthoryear{{Shore}, {Kenyon}, {Starrfield}  \&
  {Sonneborn}}{{Shore} et~al.}{1996}]{1996Shore}
{Shore} S.~N.,  {Kenyon} S.~J.,  {Starrfield} S.,   {Sonneborn} G.,  1996,
  \mn@doi [\apj] {10.1086/176692}, \href
  {https://ui.adsabs.harvard.edu/abs/1996ApJ...456..717S} {456, 717}

\bibitem[\protect\citeauthoryear{{Shore} et~al.,}{{Shore}
  et~al.}{2003}]{2003AJ....125.1507S}
{Shore} S.~N.,  et~al., 2003, \mn@doi [\aj] {10.1086/367803}, \href
  {https://ui.adsabs.harvard.edu/abs/2003AJ....125.1507S} {125, 1507}

\bibitem[\protect\citeauthoryear{{Shore} et~al.,}{{Shore}
  et~al.}{2016}]{2016A&AShore}
{Shore} S.~N.,  et~al., 2016, \mn@doi [\aap] {10.1051/0004-6361/201527856},
  \href {https://ui.adsabs.harvard.edu/abs/2016A&A...590A.123S} {590, A123}

\bibitem[\protect\citeauthoryear{{Shore} et~al.,}{{Shore}
  et~al.}{2021}]{2021ATelARAS}
{Shore} S.~N.,  et~al., 2021, The Astronomer's Telegram, \href
  {https://ui.adsabs.harvard.edu/abs/2021ATel14868....1S} {14868, 1}

\bibitem[\protect\citeauthoryear{{Snijders}}{{Snijders}}{1987}]{1987snijders}
{Snijders} M.~A.~J.,  1987, \mn@doi [\apss] {10.1007/BF00655002}, \href
  {https://ui.adsabs.harvard.edu/abs/1987Ap&SS.130..243S} {130, 243}

\bibitem[\protect\citeauthoryear{{Sokoloski}, {Luna}, {Mukai}  \&
  {Kenyon}}{{Sokoloski} et~al.}{2006}]{2006Sokoloski}
{Sokoloski} J.~L.,  {Luna} G.~J.~M.,  {Mukai} K.,   {Kenyon} S.~J.,  2006,
  \mn@doi [\nat] {10.1038/nature04893}, \href
  {https://ui.adsabs.harvard.edu/abs/2006Natur.442..276S} {442, 276}

\bibitem[\protect\citeauthoryear{{Starrfield}, {Sparks}  \&
  {Truran}}{{Starrfield} et~al.}{1974}]{1974ApJSStarrfield}
{Starrfield} S.,  {Sparks} W.~M.,   {Truran} J.~W.,  1974, \mn@doi [\apjs]
  {10.1086/190317}, \href
  {https://ui.adsabs.harvard.edu/abs/1974ApJS...28..247S} {28, 247}

\bibitem[\protect\citeauthoryear{{Starrfield}, {Truran}, {Wiescher}  \&
  {Sparks}}{{Starrfield} et~al.}{1998}]{1998MNRASStarrfield}
{Starrfield} S.,  {Truran} J.~W.,  {Wiescher} M.~C.,   {Sparks} W.~M.,  1998,
  \mn@doi [\mnras] {10.1046/j.1365-8711.1998.01312.x}, \href
  {https://ui.adsabs.harvard.edu/abs/1998MNRAS.296..502S} {296, 502}

\bibitem[\protect\citeauthoryear{{Starrfield}, {Iliadis}, {Hix}, {Timmes}  \&
  {Sparks}}{{Starrfield} et~al.}{2009}]{2009ApJStarrfield}
{Starrfield} S.,  {Iliadis} C.,  {Hix} W.~R.,  {Timmes} F.~X.,   {Sparks}
  W.~M.,  2009, \mn@doi [\apj] {10.1088/0004-637X/692/2/1532}, \href
  {https://ui.adsabs.harvard.edu/abs/2009ApJ...692.1532S} {692, 1532}

\bibitem[\protect\citeauthoryear{{Starrfield}, {Bose}, {Iliadis}, {Hix},
  {Woodward}  \& {Wagner}}{{Starrfield} et~al.}{2020}]{2020ApJStarrfield}
{Starrfield} S.,  {Bose} M.,  {Iliadis} C.,  {Hix} W.~R.,  {Woodward} C.~E.,
  {Wagner} R.~M.,  2020, \mn@doi [\apj] {10.3847/1538-4357/ab8d23}, \href
  {https://ui.adsabs.harvard.edu/abs/2020ApJ...895...70S} {895, 70}

\bibitem[\protect\citeauthoryear{{Steffen}, {Koning}, {Wenger}, {Morisset}  \&
  {Magnor}}{{Steffen} et~al.}{2011}]{2011Shape}
{Steffen} W.,  {Koning} N.,  {Wenger} S.,  {Morisset} C.,   {Magnor} M.,  2011,
  \mn@doi [IEEE Transactions on Visualization and Computer Graphics]
  {10.1109/TVCG.2010.62}, \href
  {https://ui.adsabs.harvard.edu/abs/2011ITVCG..17..454S} {17, 454}

\bibitem[\protect\citeauthoryear{{Teyssier}}{{Teyssier}}{2019}]{2019CoSka..49..217T}
{Teyssier} F.,  2019, Contributions of the Astronomical Observatory Skalnate
  Pleso, \href {https://ui.adsabs.harvard.edu/abs/2019CoSka..49..217T} {49,
  217}

\bibitem[\protect\citeauthoryear{{Vanlandingham}, {Starrfield}  \&
  {Shore}}{{Vanlandingham} et~al.}{1997}]{1997MNRAS.290...87V}
{Vanlandingham} K.~M.,  {Starrfield} S.,   {Shore} S.~N.,  1997, \mn@doi
  [\mnras] {10.1093/mnras/290.1.87}, \href
  {https://ui.adsabs.harvard.edu/abs/1997MNRAS.290...87V} {290, 87}

\bibitem[\protect\citeauthoryear{{Vanlandingham}, {Schwarz}, {Shore},
  {Starrfield}  \& {Wagner}}{{Vanlandingham}
  et~al.}{2005}]{2005ApJ...624..914V}
{Vanlandingham} K.~M.,  {Schwarz} G.~J.,  {Shore} S.~N.,  {Starrfield} S.,
  {Wagner} R.~M.,  2005, \mn@doi [\apj] {10.1086/428895}, \href
  {https://ui.adsabs.harvard.edu/abs/2005ApJ...624..914V} {624, 914}

\bibitem[\protect\citeauthoryear{{Walder}, {Folini}  \& {Shore}}{{Walder}
  et~al.}{2008}]{2008A&AWalder}
{Walder} R.,  {Folini} D.,   {Shore} S.~N.,  2008, \mn@doi [\aap]
  {10.1051/0004-6361:200809703}, \href
  {https://ui.adsabs.harvard.edu/abs/2008A&A...484L...9W} {484, L9}

\bibitem[\protect\citeauthoryear{{Woodward}, {Evans}, {Banerjee}, {Starrfield},
  {Wagner}  \& {Gehrz}}{{Woodward} et~al.}{2021}]{2021ATelIR}
{Woodward} C.~E.,  {Evans} A.,  {Banerjee} D.~P.~K.,  {Starrfield} S.,
  {Wagner} R.~M.,   {Gehrz} R.~D.,  2021, The Astronomer's Telegram, \href
  {https://ui.adsabs.harvard.edu/abs/2021ATel14866....1W} {14866, 1}

\bibitem[\protect\citeauthoryear{{della Valle} \& {Livio}}{{della Valle} \&
  {Livio}}{1995}]{1995MMRD}
{della Valle} M.,  {Livio} M.,  1995, \mn@doi [\apj] {10.1086/176342}, \href
  {https://ui.adsabs.harvard.edu/abs/1995ApJ...452..704D} {452, 704}

\makeatother
\end{thebibliography}

\bsp	
\label{lastpage}
\end{document}